\documentstyle[12pt]{livrev}
\def\lapp{\ifmmode\stackrel{<}{_{\sim}}\else$\stackrel{<}{_{\sim}}$\fi}
\def\gapp{\ifmmode\stackrel{>}{_{\sim}}\else$\stackrel{>}{_{\sim}}$\fi}
\begin{document}
\title{Binary and Millisecond Pulsars at the New Millennium}

\author{D. R. Lorimer ({\it dunc@naic.edu})\\
Arecibo Observatory, HC3 Box 53995,\\
Puerto Rico 00612, USA}
\date{To appear in: {\it Living Reviews in Relativity} (www.livingreviews.org)}
\maketitle

\begin{center} {\large \bf Abstract} \end{center}
\noindent
We review the properties and applications of binary and millisecond
pulsars. Our knowledge of these exciting objects has greatly increased
in recent years, mainly due to successful surveys which have brought
the known pulsar population to over 1300. There are now 56 binary and
millisecond pulsars in the Galactic disk and a further 47 in globular
clusters. This review is concerned primarily with the results and spin-offs from these surveys which are of particular interest to the relativity community. 

\section{Preamble}
\label{sec:preamble}

In the 34 years that have elapsed since the discovery \cite{hbp+68} of pulsars,
rapidly rotating highly magnetised neutron stars, the study of these
fascinating objects has resulted in many applications in physics and
astronomy. Striking examples include the confirmation of the
existence of gravitational radiation \cite{nobpr1993} as predicted by
general relativity \cite{tw82,tw89} and the first
detection of an extra-solar planetary system \cite{wf92,psrplanets}. 

\begin{figure}[hbt]
\setlength{\unitlength}{1in}
\begin{picture}(0,2.75)
\put(0.5,3.5){\includegraphics{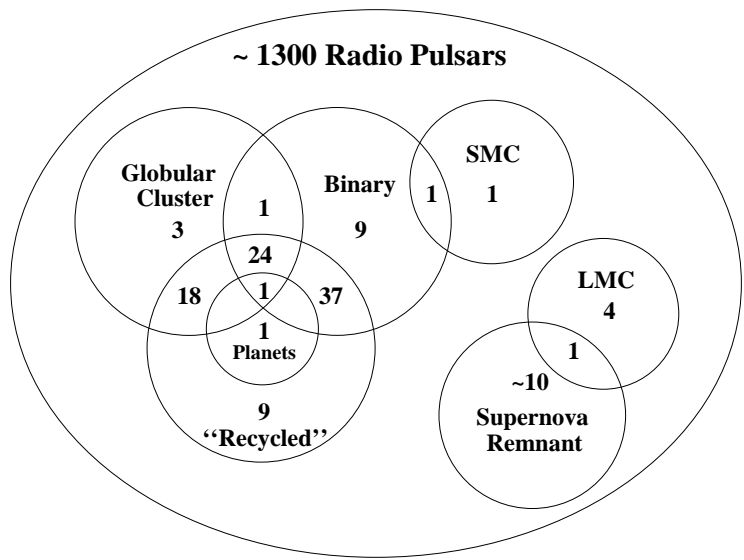}}
\end{picture}
\caption[]
{\sl
The numbers and locations of the various types of radio pulsars
known as of December 2000.  The large and small Magellanic
clouds are denoted by LMC and SMC.
}
\label{fig:venn}
\end{figure}
The diverse zoo of radio pulsars currently known is summarized
graphically by the Venn diagram in Fig.~\ref{fig:venn}.  Many new
binary systems containing neutron stars are being discovered as a
result of the latest generation of pulsar surveys. This review is
concerned primarily with the results and spin-offs from these
surveys which are of particular interest to the relativity community.

\subsection*{What's new in this review?}

Since the first version of this article was written back in 1997/8
\cite{lor98e} a number of pulsar surveys using the Parkes radio 
telescope \cite{man01} have discovered almost 700 pulsars. As a
result, the sample size is now double what it was in 1997.  Many of
the exciting new discoveries from these searches are discussed in this
review. Up-to-date tables of parameters of binary and millisecond
pulsars are included as an appendix. Several new sections/figures have
been added and existing sections reworked and modularized to make the
review more self-contained and (hopefully!)  easier to read in an html
setting.  We begin in \S \ref{sec:intro} with an overview of the
pulsar phenomenon, the key observed population properties, the origin
and evolution of pulsars and an introduction to pulsar search
techniques. In \S \ref{sec:gal}, we review present understanding in
pulsar demography, discussing selection effects and the techniques
used to correct for them in the observed sample. This leads to robust
estimates of the total number of normal and millisecond pulsars (\S
\ref{sec:nmsppop}) and relativistic binaries (\S \ref{sec:relpop}) in
the Galaxy and has implications for the detection of gravitational
radiation from these systems. We discuss pulsar timing in \S
\ref{sec:pultim}. One application of these exceptional clocks, a 
sensitive detector of long-period gravitational waves, is discussed 
in \S \ref{sec:gwdet}.  We conclude with a brief outlook to the future 
in \S \ref{sec:future}.

\section{An Introduction to Pulsar Astronomy}
\label{sec:intro}

Many of the basic observational facts about radio pulsars were
established shortly after their discovery \cite{hbp+68} by Bell and
Hewish in 1967.  In the intervening years, theoretical and
observational progress has flourished. Although there are many
remaining questions, particularly about the emission mechanism, the
basic model has long been established beyond all reasonable doubt {\it
viz:} Pulsars are rapidly rotating, highly magnetised neutron stars
formed during the supernova explosions of massive ($\sim$ 5--10
M$_{\odot}$) stars.  In the following, we discuss the basic
observational properties most relevant to this review.

\subsection{The lighthouse model}
\label{sec:light}

\begin{figure}[hbt]
\setlength{\unitlength}{1in}
\begin{picture}(0.0,2.1)
\put(-0.6,-4.2){\includegraphics{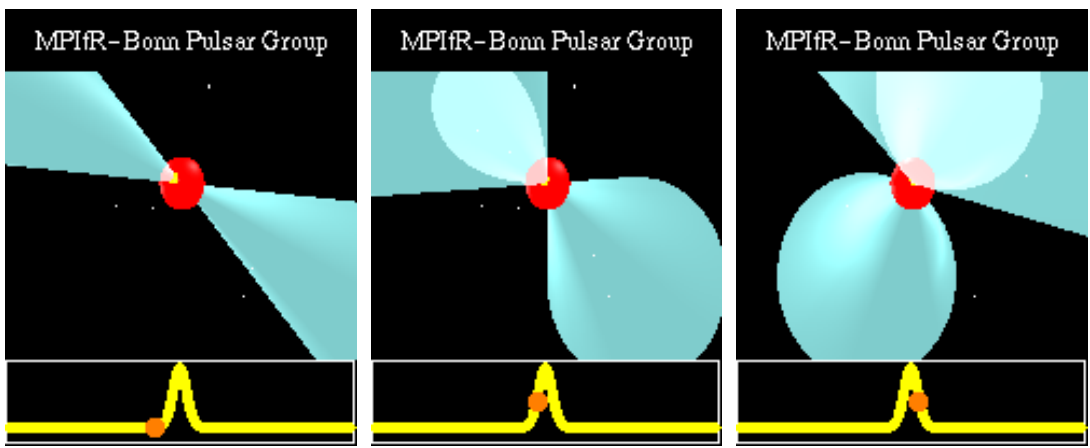}}
\end{picture}
\caption[]
{\sl
The rotating neutron star (or ``lighthouse'') model for pulsar emission.  
Click here to see the movie in action. Animation designed by Michael Kramer.
}
\label{fig:rotns}
\end{figure}

An animation showing the rotating neutron star or ``lighthouse model'' 
of the basic pulsar phenomenon is shown in Fig.~\ref{fig:rotns}.
%
%
As the neutron star spins, charged particles are accelerated out along
magnetic field lines in the magnetosphere (depicted by the light blue
cones). This acceleration causes the particles to emit electromagnetic
radiation, most readily detected at radio frequencies as a sequence of
observed pulses produced as the magnetic axis (and hence the radiation
beam) crosses the observer's line of sight each rotation. The
repetition period of the pulses is therefore simply the rotation
period of the neutron star.  The moving ``tracker ball'' on the
pulse profile in the animation shows the relationship
between observed intensity and rotational phase of the neutron star.

Neutron stars are extremely stable rotators. They are essentially
large celestial flywheels with moments of inertia $\sim
10^{45}$ g cm$^2$.  The rotating neutron star model, independently 
developed by Pacini and Gold in 1968 \cite{pac68,gol68}, 
predicts a gradual increase in the pulse period as the
outgoing radiation carries away rotational kinetic energy.  This model
became universally accepted when a period increase of 36.5 ns per day
was measured for the pulsar in the Crab nebula \cite{rc69b}, enabling
Gold \cite{gol69} to show that a rotating neutron star with a large
magnetic field must be the dominant energy supply for the nebula.

\subsection{Pulse profiles}
\label{sec:profs}

Pulsars are weak radio sources. Measured flux densities, usually quoted
in the literature for a radio frequency of 400 MHz, vary between 0.1 and
5000 mJy (1 Jy $\equiv 10^{-26}$ W m$^{-2}$ Hz$^{-1}$).  This means
that, even with a large radio telescope, the coherent addition of many
thousands of pulses is required in order to produce an integrated
profile.  Remarkably, although the individual pulses vary dramatically 
from pulse to pulse, at any particular observing frequency the integrated 
profile is very stable. The pulse profile can thus be thought of as a 
fingerprint of the emission beam.
%
%
%
\begin{figure}[hbt]
\setlength{\unitlength}{1in}
\begin{picture}(0,2)
\put(0.8,-2.1){\includegraphics{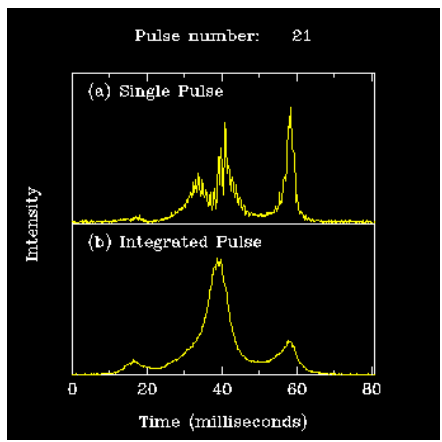}}
\end{picture}
\caption[]
{\sl
Single pulses from PSR B0329+54. Click here to see the movie in action.
}
\label{fig:sp}
\end{figure}
The animation in Fig.~\ref{fig:sp} shows a sequence of consecutive
single pulses from PSR B0329+54\footnote{Pulsars are named with a PSR
prefix followed by a ``B'' or a ``J'' and their celestial
coordinates. Those pulsars discovered prior to 1990 are usually
referred to by their ``B'' names (Besselian 1950 system).  More recent
discoveries are only referred to by their ``J'' names (Julian 2000
system).}  one of the brightest pulsars.  This pulsar is seen in the
animation to stabilise into its characteristic 3-component form after
the summation of a number of seemingly erratic single pulses. 
Stabilization time-scales are typically several hundred pulses
\cite{hmt75}.  This property is of key importance in pulsar timing
measurements discussed in detail in \S \ref{sec:pultim}.

\begin{figure}[hbt]
\setlength{\unitlength}{1in}
\begin{picture}(0,3.25)
\put(0.5,3.4){\includegraphics{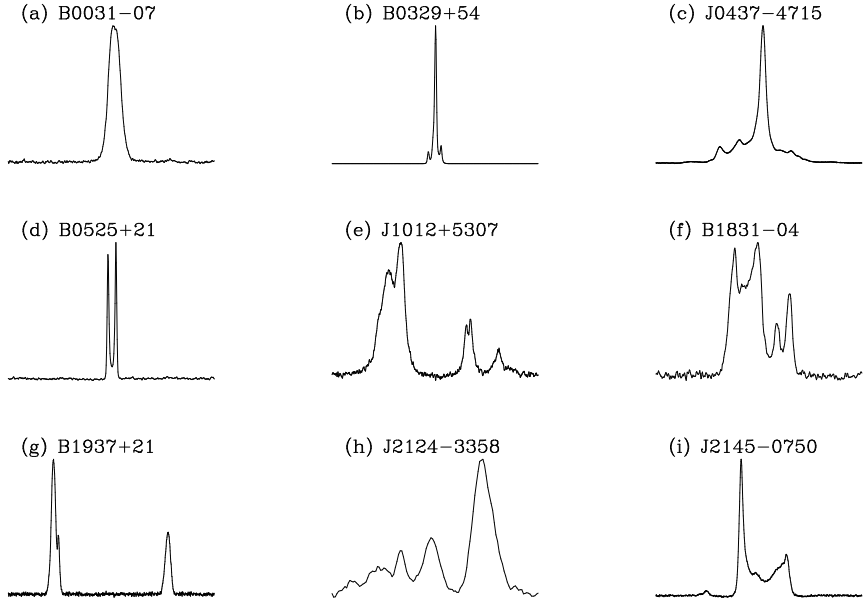}}
\end{picture}
\caption[]
{\sl
A variety of integrated pulse profiles taken from the available
literature. References: (a,b,d,f: \cite{gl98}); (c: \cite{bjb+97});
(e,g,i: \cite{kxl+98}); (h: \cite{bbm+97}). Each profile represents
360 degrees of rotational phase. These profiles are part of a database
of over 2600 multi-frequency pulse profiles for over 600 pulsars that
is available on-line \cite{epndb}.
}
\label{fig:profs}
\end{figure}

Fig.~\ref{fig:profs} shows the rich diversity in
morphology from simple single-component profiles to examples in which
emission is observed over the entire pulse.  The astute reader will
notice two examples of ``interpulses'' --- a secondary pulse separated
by about 180 degrees from the main pulse. The most natural
interpretation for this phenomenon is that the two pulses originate
from opposite magnetic poles of the neutron star (see however
\cite{ml77}). Since this is an unlikely viewing angle we would expect
interpulses to be a rare phenomenon. Indeed this is the case: the
fraction of known pulsars in which interpulses are observed in their
pulse profiles is only a few percent.

\begin{figure}[hbt]
\setlength{\unitlength}{1in}
\begin{picture}(0,1.5)
\put(0.75,0){\includegraphics{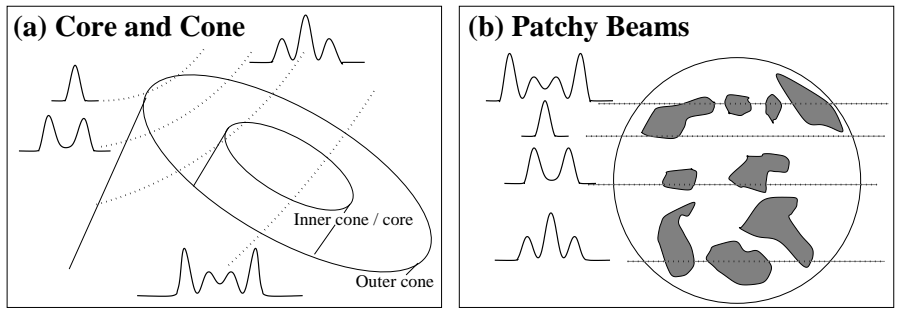}}
\end{picture}
\caption[]
{\sl
Phenomenological models for pulse shape morphology produced
by different line-of-sight cuts of the beam. (Figure designed
by M.~Kramer and A.~von Hoensbroech).
}
\label{fig:shapes}
\end{figure}

Two contrasting phenomenological models to explain the observed pulse
shapes are shown in Fig.~\ref{fig:shapes}. The ``core and cone''
model, proposed by Rankin \cite{ran83}, depicts the beam as a core
surrounded by a series of nested cones. Alternatively, the ``patchy
beam'' model, championed by Lyne and Manchester \cite{lm88,hm01}, has
the beam populated by a series of randomly-distributed emitting
regions.  Further work in this area, particularly in trying to
quantify the variety of pulse shapes (number of distinct components
and the relative fraction that they occur) is necessary to improve our
understanding of the fraction of sky covered by the radio pulsar
emission beam. We return to this topic in the context of pulsar
demography later on in \S \ref{sec:corsamp}.

\subsection{The pulsar distance scale}
\label{sec:dist}

\begin{figure}[hbt]
\setlength{\unitlength}{1in}
\begin{picture}(0,2.2)
\put(0.2,3.45){\includegraphics{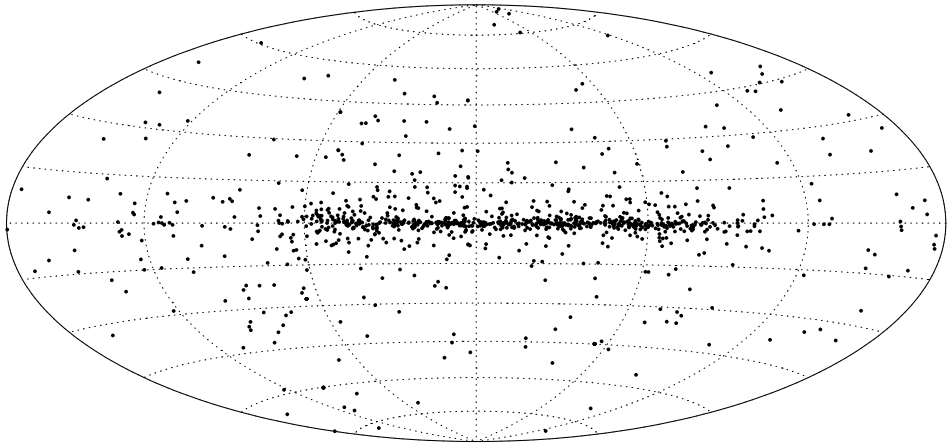}}
\end{picture}
\caption[]
{\sl
The sky distribution of 1026 pulsars in Galactic coordinates. The plane of
the Galaxy is the central horizontal line. The Galactic centre
is the midpoint of this line.
}
\label{fig:aitoff}
\end{figure}

From the sky distribution shown in Fig.~\ref{fig:aitoff} it is immediately 
apparent that pulsars are strongly concentrated along the Galactic plane. 
This indicates that pulsars populate the disk of our Galaxy. Unlike most
other classes of astrophysical objects, quantitative estimates of the distances
to each pulsar can be made from an effect known as {\it pulse
dispersion}, the delay in pulse arrival times across a finite bandwidth.
Dispersion occurs because the group velocity of the pulsed
radiation through the ionised component of the 
interstellar medium is frequency
dependent: pulses emitted at higher radio frequencies travel faster
through the interstellar medium, arriving earlier than those emitted
at lower frequencies.  The delay $\Delta t$ in arrival times between a
high frequency $\nu_{\rm hi}$ and a low frequency $\nu_{\rm lo}$
pulse is given \cite{ls98} by
\begin{equation}
\label{equ:defdt}
 \Delta t = 4150 {\rm s} \, \, \times 
 (\nu_{\rm lo}^{-2} - \nu_{\rm hi}^{-2})  \times {\rm DM},
\end{equation}
where the frequencies are in MHz and
the dispersion measure DM (cm$^{-3}$ pc) is the integrated
column density of free electrons along the line of sight:
\begin{equation}
\label{equ:defdm}
{\rm DM} = \int_{\rm 0}^{d} \,\, n_{\rm e} \,\, dl.
\end{equation}
Here, $d$ is the distance to the pulsar (pc) and $n_{\rm e}$ is the
free electron density (cm$^{-3}$).  From equation \ref{equ:defdm}
it is obvious that a measurement of the delay
across a finite bandwidth yields the DM.  Pulsars at large distances
have higher column densities and therefore larger DMs than those
pulsars closer to Earth so that, from Equation \ref{equ:defdt}, the
dispersive delay across the bandwidth is greater.  Hence, given the
DM, the distance can be estimated from a model of the Galactic
distribution of free electrons. 

The electron density model is calibrated from the pulsars with
independent distance estimates and measurements of scattering
for lines of sight towards various Galactic and extragalactic sources.
Independent distance estimates now exist for over 100 pulsars based on
three basic techniques: neutral
hydrogen absorption, trigonometric parallax (measured either with an
interferometer or through pulse time-of-arrival techniques) and from
associations with objects of known distance ({\it i.e.}~supernova
remnants, globular clusters and the Magellanic Clouds). 
Based on these data, Taylor \& Cordes \cite{tc93} have 
developed an electron density model which is free from large
systematic trends and can be used to provide distance estimates with
an uncertainty of $\sim$ 30\%.  However, use of this model to estimate
distances to {\it individual pulsars} may result in uncertainties by
as much as a factor of two. This model is currently being refined
following recent pulsar discoveries and independent distance and
scattering measurements (J.~Cordes, private communication).

\subsection{Normal and millisecond pulsars}
\label{sec:nms}

\subsubsection{Spin parameters}
\label{sec:spinpars}

The present public-domain catalogue, available on-line at Princeton
\cite{pripsr}, contains up-to-date parameters for $706$ pulsars.
Parameters for many of the new Parkes multibeam pulsar surveys are
also available on-line \cite{bmc+00,mbeampsr}.  Most of these are
``normal'' in the sense that their pulse periods $P$ are of order one
second and are observed to increase secularly at rates $\dot{P}$ of
typically $10^{-15}$ s/s. A growing fraction of the observed sample
are the so--called ``millisecond pulsars'', which have spin periods
primarily in the range 1.5 and 30 ms and rates of slowdown $\lapp
10^{-19}$ s/s. The first millisecond pulsar discovered, B1937+21
\cite{bkh+82}, with $P=1.5578$ ms, remains the most rapidly rotating
neutron star presently known.

\begin{figure}[hbt]
\setlength{\unitlength}{1in}
\begin{picture}(0,3)
\put(1.3,3.6){\includegraphics{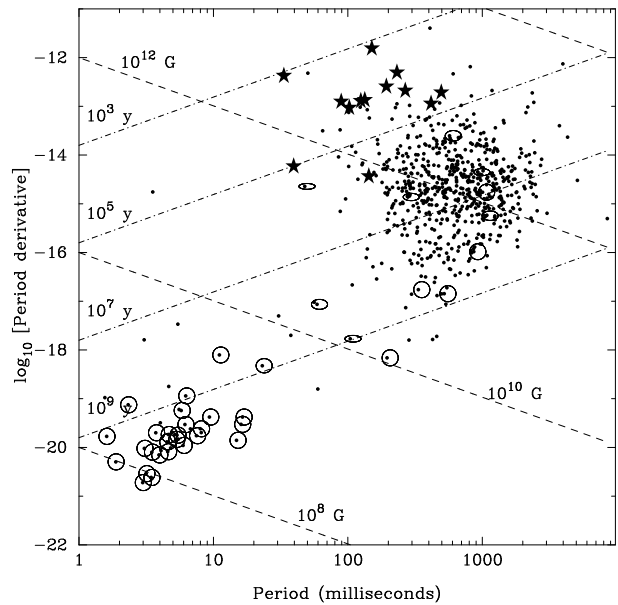}}
\end{picture}
\caption[]
{\sl
The ubiquitous $P - \dot{P}$ diagram shown for a sample of radio
pulsars.  Those objects known to be members of binary systems are
highlighted by a circle (for low-eccentricity orbits) or an ellipse
(for elliptical orbits). Pulsars thought to be associated with
supernova remnants are highlighted by the starred symbols.
}
\label{fig:ppdot}
\end{figure}

A very useful means of demonstrating the distinction between these two
classes is the ``$P$--$\dot{P}$ diagram'' -- a logarithmic scatter
plot of the observed pulse period versus the period derivative. 
As shown in Fig.~\ref{fig:ppdot}, normal pulsars occupy the
majority of the upper right hand part of the diagram, while the
millisecond pulsars reside in the lower left hand part of the diagram.
The differences in $P$ and $\dot{P}$ imply different ages and surface
magnetic field strengths. By treating the pulsar as a rotating
magnetic dipole, one may show that the surface magnetic field strength
is proportional to $(P \dot{P})^{1/2}$ \cite{mt77}.  Lines of constant
magnetic field strength are drawn on Fig.~\ref{fig:ppdot}, together
with lines of constant {\it characteristic age} ($\tau_c =
P/(2\dot{P})$).  Typical inferred magnetic fields and
ages are $10^{12}$ G and $10^{7}$ yr for the normal pulsars and
$10^{8}$ G and $10^{9}$ yr for the millisecond pulsars.

\subsubsection{Binary companions}
\label{sec:bincomps}

As can be inferred from Fig.~\ref{fig:venn}, just under 4\% of 
all known pulsars in the Galactic disk are members of binary
systems.  Timing measurements (\S \ref{sec:pultim}) place useful
constraints on the masses of the companions which, supplemented by
observations at other wavelengths, tell us a great deal about their
nature. The present sample of orbiting companions are either white
dwarfs, main sequence stars, or other neutron stars.  Two notable
hybrid systems are the ``planet pulsars'' PSR B1257+12 and
B1620--26. B1257+12 is a 6.2-ms pulsar accompanied by at least three
Earth-mass bodies \cite{wf92,psrplanets,wdk+00} while B1620--26, an 
11-ms pulsar in the globular cluster M4, is part of a triple system
with a white dwarf and a high-mass planet \cite{tat93,bfs93,tacl99}.

\begin{figure}[hbt]
\setlength{\unitlength}{1in}
\begin{picture}(0,2.8)
\put(0.5,3.45){\includegraphics{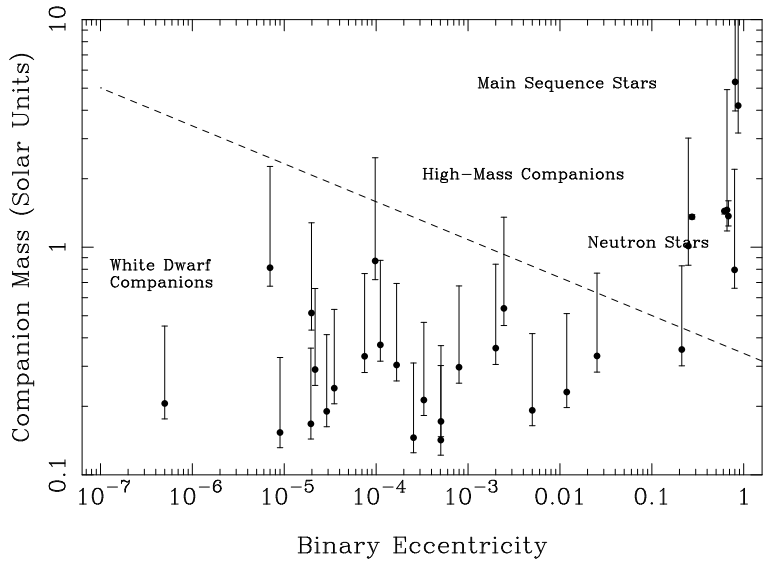}}
\end{picture}
\caption[]
{\sl
Companion mass versus orbital eccentricity for the sample of binary pulsars.
}
\label{fig:em}
\end{figure}

A very important additional difference between normal and millisecond
pulsars is the presence of an orbiting companion.  Orbital companions
are much more commonly observed around millisecond pulsars ($\sim
80$\% of the observed sample) than around the normal pulsars ($\lapp
1$\%).  Fig.~\ref{fig:em} is a scatter plot of orbital eccentricity
versus mass of the companion. The dashed line serves merely to guide
the eye in this figure. Binary systems lying below the line are those
with low-mass companions ($\lapp 0.7$ M$_{\odot}$ --- predominantly
white dwarfs) and essentially circular orbits: $10^{-5} \lapp e \lapp
0.1$.  Binary pulsars with high-mass companions ($\gapp 1$ M$_{\odot}$
--- neutron stars or main sequence stars) are in eccentric orbits:
$0.15 \gapp e \gapp 0.9$ and lie above the line.

\subsubsection{Evolutionary scenarios}
\label{sec:evolution}

\begin{figure}[hbt]
\setlength{\unitlength}{1in}
\begin{picture}(0,4.3)
\put(1.2,0.0){\includegraphics{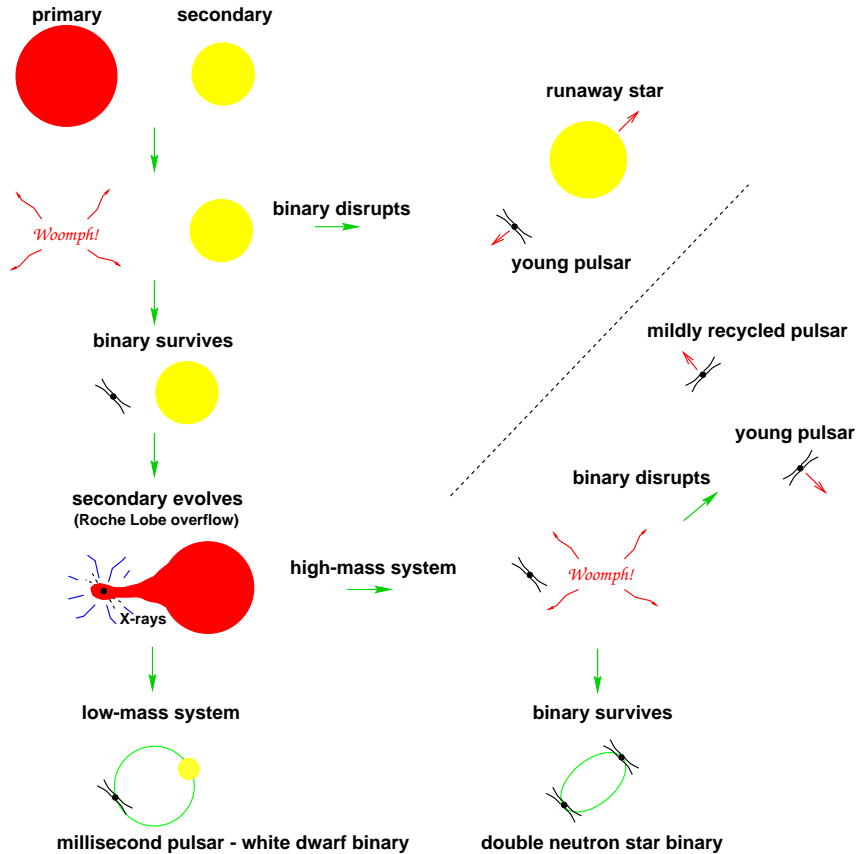}}
\end{picture}
\caption[]{\sl
Cartoon showing various evolutionary scenarios involving binary pulsars.
}
\label{fig:bevol}
\end{figure}

The presently favoured model to explain the formation of the various
types of systems has been developed over the years by a number of
authors \cite{bk74,fv75,sb76,acrs82}. The model is
sketched in Fig.~\ref{fig:bevol} and is now qualitatively summarised.

Starting with a binary star system, a neutron star is formed during
the supernova explosion of the initially more massive star which has
an inherently shorter main sequence lifetime. From the virial theorem
it follows that the binary system gets disrupted if more
than half the total pre-supernova mass is ejected from the system
during the explosion \cite{hil83,bv91}.  
In addition, the fraction of surviving binaries
is affected by the magnitude and direction of any impulsive ``kick''
velocity the neutron star receives at birth \cite{hil83,bai89}.  Those
binary systems that disrupt produce a high-velocity isolated neutron
star and an OB runaway star \cite{bla61}.  
The high binary disruption probability during the explosion
explains, qualitatively at least, why so few normal pulsars have companions.
Over the next $10^{7}$ yr
or so after the explosion, the neutron star may be observable as a
normal radio pulsar spinning down to a period $\gapp$ several
seconds. After this time, the energy output of the star
diminuishes to a point where it no longer produces significant
radio emission.

For those few binaries that remain bound, and in which the companion
is sufficiently massive to evolve into a giant and overflow its Roche
lobe, the old spun-down neutron star can gain a new lease of life as a
pulsar by accreting matter and therefore angular momentum at the expense
of the orbital angular momentum of the binary system \cite{acrs82}. 
The term ``recycled pulsar'' is often used to
describe such objects. During this accretion phase, the X-rays
produced by the liberation of gravitational energy of the infalling matter
onto the neutron star mean
that such a system is expected to be visible as an X-ray binary
system. Two classes of X-ray binaries relevant to binary and
millisecond pulsars exist, {\it viz:} neutron stars with high-mass or
low-mass companions. For a detailed review of the X-ray binary
population, including systems likely to contain black holes rather
than neutron stars, the interested reader is referred to \cite{bv91}.

The high-mass companions are massive enough to explode as a supernova,
producing a second neutron star. If the binary system is lucky enough
to survive {\it this} explosion, it ends up as a double neutron star
binary. The classic example is PSR B1913+16 \cite{ht75a}, a 59-ms 
radio pulsar with a characteristic age of $\sim 10^8$ yr which orbits
its companion every 7.75 hr \cite{tw82,tw89}. In this formation scenario, 
PSR B1913+16 is an example of the older, first-born, neutron star that has
subsequently accreted matter from its companion. So far there are
no clear examples of systems where the second-born neutron star
is observed as a radio pulsar. In the case of {PSR B1820$-$11 \cite{lm89}, which may be an example, the mass of the companion is not well
determined, so either a main-sequence \cite{pv91} or a
white dwarf companion \cite{py99} are plausible alternatives.
This lack of observation of second-born neutron stars as radio pulsars is probably reasonable 
when one realises that the observable
lifetimes of recycled pulsars are much larger than those of normal
pulsars.  As discussed in \S \ref{sec:nsns}, double neutron star
binary systems are very rare in the Galaxy --- another indication that
the majority of binary systems get disrupted when one of the
components explodes as a supernova.  Systems disrupted after the
supernova of the secondary form a mildly-recycled isolated pulsar and
a young pulsar formed during the explosion of the secondary.

Although no system has so far been found in which both neutron stars
are visible as radio pulsars, timing measurements of three systems
show that the companion masses are 1.4 M$_{\odot}$ --- as 
expected for neutron stars \cite{st83}. In addition, no optical
counterparts are seen. Thus, we conclude that these unseen companions
are neutron stars that are either too weak to be detected, no longer
active as radio pulsars, or their emission beams do not intersect our
line of sight. The two known young radio pulsars with main sequence
companions massive enough to explode as a supernova probably represent
the intermediate phase between high-mass X-ray binaries and double
neutron star systems \cite{jml+92,kjb+94}.

The companions in the low-mass X-ray binaries evolve
and transfer matter onto the neutron star on a much longer time-scale,
spinning it up to periods as short as a few ms \cite{acrs82}. This
model has gained strong support in recent years from the discoveries
of quasi-periodic kHz oscillations in a number of low-mass X-ray
binaries \cite{wz97}, as well as Doppler-shifted 2.49-ms X-ray
pulsations from the transient X-ray burster SAX J1808.4--3658
\cite{wv98,cm98}.  At the end of the spin-up phase, the secondary
sheds its outer layers to become a white dwarf in orbit around a
rapidly spinning millisecond pulsar. Presently $\sim 10$ of these
systems have compelling optical identifications of the white dwarf
companion \cite{bbb93,bkb+95,lcf+96,lfc96}.  Perhaps the best example
is the white dwarf companion to the 5.25-ms pulsar J1012+5307
\cite{nll+95,lfln95}. This 19$^{\rm th}$ magnitude white dwarf is bright
enough to allow measurements of its surface gravity and orbital
velocity \cite{vbk96}. 

The range of white dwarf masses observed is becoming broader. Since
this article originally appeared in 1998 the number of 
``intermediate-mass binary pulsars'' \cite{cam96c} 
has grown significantly \cite{clm+01}. These systems are distinct to the
``classical'' millisecond pulsar--white dwarf binaries like PSR
J1012+5307 in several ways: (1) the spin period of the radio pulsar
is generally longer (9--200 ms); (2) the mass of the white dwarf is
larger (typically close to 1 M$_{\odot}$); (3) the orbital
eccentricity, while still essentially circular, is often
significantly larger ($\sim 10^{-3}$). It is not presently clear
whether these systems originated from either low- or high-mass X-ray
binaries. It was suggested by van den Heuvel \cite{vdh94} that they have
more in common with high-mass systems, the difference being that the
secondary star was not sufficiently massive to explode as a
supernova. Instead it formed a white dwarf. Detailed studies of
this sub-population of binary pulsars are required for further 
understanding in this area.

Another relatively poorly understood area is the existence of 
solitary millisecond pulsars in the Galactic disk (which comprise 
just under 20\% of all Galactic millisecond pulsars). Although
it has been proposed that the millisecond pulsars have got rid
of their companion by ablation, as appears to be happening in
the PSR B1957+20 system \cite{fst88}, it is not clear whether
the time-scales for this process are feasible. There is some
observational evidence that suggests that solitary millisecond
pulsars are less luminous than binary millisecond pulsars
\cite{bjb+97,kxl+98}. If  confirmed by future discoveries, this
would need to be explained by any viable evolutionary model.

\subsubsection{Space velocities}
\label{sec:pvel}

Pulsars have long been known to have space velocities at least an order 
of magnitude larger than those of their main sequence progenitors, which 
have typical values between 10 and 50 km s$^{-1}$. The first direct 
evidence for large velocities came from optical observations of the Crab 
pulsar in 1968 \cite{tri68}, showing that the neutron star has a velocity 
in excess of 100 km s$^{-1}$. Proper motions for about 100 pulsars have 
subsequently been measured largely by radio interferometric techniques 
\cite{las82,bmk+90b,fgl+92,hla93}. These data imply a broad velocity 
spectrum ranging from 0 to over 1000 km s$^{-1}$ \cite{ll94}.

%
%
\begin{figure}[hbt]
\setlength{\unitlength}{1in}
\begin{picture}(0,2.35)
\put(-0.3,-3.8){\includegraphics{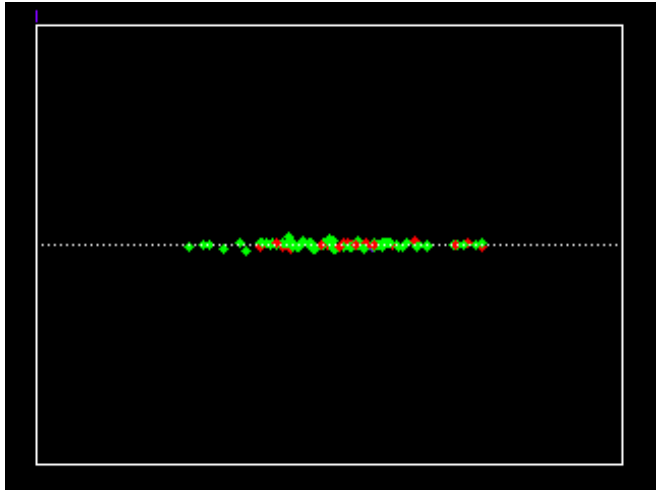}}
\end{picture}
\caption[]
{\sl
A simulation following the motion of 100 pulsars in a model
gravitational potential of our Galaxy for 200 Myr.  The view is
edge-on {\it i.e.}~the horizontal axis represents the Galactic plane
(30 kpc across) while the vertical axis represents $\pm 10$ kpc from
the plane. This snapshot shows the initial configuration of young
neutron stars.  Click here to see the movie in action.
}
\label{fig:migrate}
\end{figure}

Such large velocities are perhaps not surprising, given the violent
conditions under which neutron stars are formed. Shklovskii \cite{shk70}
demonstrated that, if the explosion is only slightly asymmetric, an
impulsive ``kick'' velocity of up to 1000 km s$^{-1}$ is imparted to
the neutron star. In addition, if the neutron star progenitor was 
a member of a binary system prior to the explosion, the pre-supernova
orbital velocity will also contribute to the resulting speed of
the newly-formed pulsar. High-velocity pulsars born close to the
Galactic plane quickly migrate to higher Galactic latitudes. This 
migration is seen in Fig.~\ref{fig:migrate}, a dynamical simulation of
the orbits of 100 neutron stars in a model of the Galactic
gravitational potential. Given such a broad velocity spectrum, as
much as half of all pulsars will eventually escape the gravitational
potential of the Galaxy and end up in intergalactic space \cite{ll94,cc98}.

Based on the proper motion data, recent studies have demonstrated that
the mean birth velocity of normal pulsars is $\sim$ 450 km s$^{-1}$
(\cite{ll94,lbh97,cc98,fbb98}; see, however, also \cite{har97,hp97}).
This is significantly larger than the velocities of millisecond and
binary pulsars. Recent studies suggest that their mean birth velocity
is likely to be in the range $\sim 80 - 140$ km s$^{-1}$
\cite{lor95,cc97,lml+98}. The main reason for this difference surely
lies in the fact that about 80\% of the millisecond pulsars are
members of binary systems (\S \ref{sec:nms}) which could not have 
survived had the neutron star received a substantial kick velocity.

\subsection{Searching for pulsars}
\label{sec:searching}

Pulsar searching is, conceptually at least, a rather simple process
--- the detection of a dispersed, periodic signal hidden in a noisy
time series collected using a large radio telescope. The search
 These data are then
analysed for periodic signals. We give here a
brief description of the basic search techniques.  More detailed
discussions can be found elsewhere \cite{lyn88,nic92,lor98}.

A schematic pulsar search is shown in Fig.~\ref{fig:search}.  The
finite bandwidth is split up into a number of channels, typically
using a filterbank or a correlator (see e.g.~\cite{bckw90}),
either of which usually provides a much finer frequency channelisation
than the eight channels shown for illustrative purposes in
Fig.~\ref{fig:search}.  The channels are then de-dispersed 
(see \S \ref{sec:dispandscatt})
to form a single noisy time series. An efficient way to
find a periodic signal in these data is to take the Fast Fourier
Transform (FFT) and plot the resulting amplitude spectrum. For a
narrow pulse the spectrum will show a family of harmonics. To detect
weaker signals still, a harmonic summing technique is usually
implemented \cite{lyn88}. The best candidates are saved and the whole
process is repeated for another trial DM.

\begin{figure}[hbt]
\setlength{\unitlength}{1in}
\begin{picture}(0,2.4)
\put(0.7,-0.1){\includegraphics{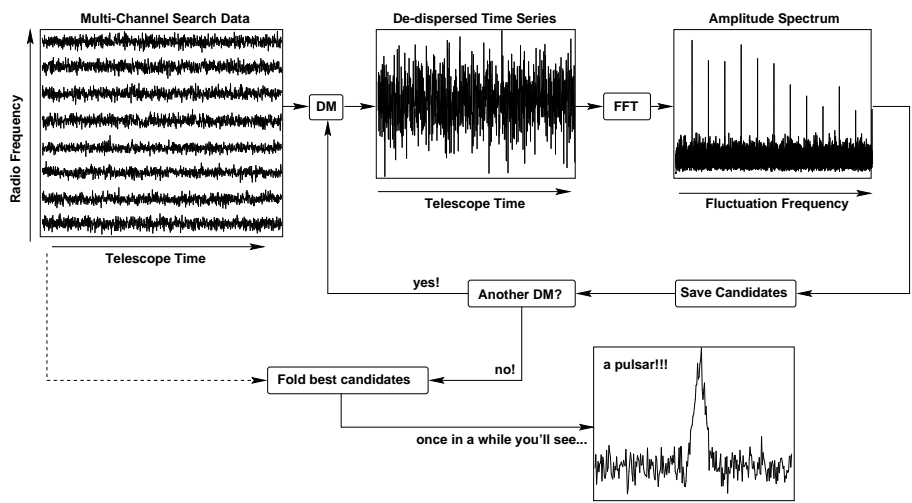}}
\end{picture}
\caption[]
{\sl
Schematic summarising the essential steps in a ``standard'' pulsar search.
}
\label{fig:search}
\end{figure}

After the data have been processed for a suitable range of DM, a 
list of pulsar candidates is compiled and the raw time series data are
folded modulo each candidate period.  In practise the analysis is
often hampered by the presence of periodic interference sources which
can often look very ``pulsar-like''. Although interference excision
schemes (usually based on coincidence analyses of data taken from
different points on the sky) work fairly well, interference is an
ever-increasing problem in radio astronomy and considerable efforts
are required to carry out sensitive searches.

\subsection{Where to look for binary and millisecond pulsars}
\label{sec:wheretolook}

Following the above discussions on demography and evolution,
it is instructive to briefly summarize the rationale behind the
major searches being carried out at the present time.

\subsubsection{All-sky searches}
\label{sec:allsky}

The oldest radio pulsars form a virialised population of
stars oscillating in the Galactic gravitational potential. The scale
height for such a population is at least 500 pc, about 10 times that
of the massive stars which populate the Galactic plane. Since the
typical ages of millisecond pulsars are several Gyr or more, we
expect, from our vantage point in the Galaxy, to be in the middle of
an essentially isotropic population of nearby sources. All-sky
searches for millisecond pulsars at high Galactic latitudes have been
very effective in probing this population. Much of the initial
interest and excitement in this area was started at Arecibo when
Wolszczan discovered two classic recycled pulsars at high latitudes:
the neutron star binary B1534+12 \cite{wol91a} and the planets pulsar
B1257+12 \cite{wf92}. Surveys carried out at Arecibo, Parkes, Jodrell Bank
and Green Bank by others in the 1990s found many other millisecond and
recycled pulsars in this way.  Camilo has written several excellent
reviews of these surveys \cite{cam95,cam97,cam98}. See also Tables 
\ref{tab:imsps}, \ref{tab:ebpsrs} and \ref{tab:bmsps} in the appendix.

\subsubsection{Searches close to the plane of our Galaxy}
\label{sec:plane}

Young pulsars are most likely to be found near to their place of
birth, close to the Galactic plane. This is the target region of one of
the Parkes multibeam surveys and has already resulted in the discovery
of around 600 new pulsars \cite{clm+00,mlc+00}, almost half the number
currently known! Such a large haul inevitably results in a number of
interesting individual objects such as: PSR J1141$-$6545, a young
pulsar in a relativistic 4-hr orbit around a white dwarf \cite{klm+00}; 
PSR J1740$-$3052, a young pulsar orbiting an $\sim 11$ M$_{\odot}$ star
(probably a giant \cite{mlc+00}); several intermediate-mass binary pulsars
\cite{clm+01} and a likely double neutron star system PSR J1811$-$1736
\cite{lcm+00}. 

\subsubsection{Searches at intermediate Galactic latitudes}
\label{sec:intlat}

In order to probe more deeply into the population of millisecond and
recycled pulsars than possible at high Galactic latitudes, Edwards et
al.~have recently completed a survey of intermediate latitudes with
the Parkes multibeam system \cite{edw00,eb01}. The results of this
survey are extremely exciting --- 58 new pulsars including 8
relatively distant recycled objects. Two of the new recycled pulsars
from this survey recently announced by Edwards \& Bailes
\cite{eb01} are mildly relativistic neutron star-white dwarf 
binaries. An analysis of the full results from this survey should
significantly improve our knowledge on the Galaxy-wide population and
birth-rate of millisecond pulsars.

\subsubsection{Targeted searches of globular clusters}
\label{sec:globs}

Globular clusters have long been known to be breeding grounds for
millisecond and binary pulsars. The main reason for this is the high
stellar density in globular clusters relative to most of the rest of
the Galaxy. As a result, low-mass X-ray binaries are almost 10 times more
abundant in clusters than in the Galactic disk. In addition, exchange
interactions between binary and multiple systems in the cluster can
result in the formation of exotic binary systems.  Since a single
globular cluster usually fits well within a single telescope beam,
deep targeted searches can be made. Once the DM of a pulsar
is known in a globular cluster, the DM parameter space for subsequent
searches is essentially fixed. This allows computation power to be
invested in so-called acceleration searches for short-period binary systems
(see \S \ref{sec:accn}).  To date, searches have revealed 47 pulsars
in globular clusters (see Table \ref{tab:gcpsrs} in the appendix for a list and the
review by Kulkarni \& Anderson \cite{ka96}). Highlights include the
double neutron star binary in M15 \cite{pakw91} and a low-mass binary
system with a 95-min orbital period in 47~Tucanae \cite{clf+00}, one
of 20 millisecond pulsars currently known in this cluster alone
\cite{fcl+01}.  On-going surveys of clusters continue to yield new
discoveries \cite{rgh+01,dlm+01}.

\subsection{Going further}

Two excellent graduate-level monographs are available: the classic 1970s
text {\it Pulsars}
by Manchester \& Taylor \cite{mt77} and the more up-to-date
{\it Pulsar Astronomy} by Lyne
\& Smith \cite{ls98}. Those wishing to approach the subject from a
more theoretical viewpoint are advised to read Michel's {\it The
Theory of Neutron Star Magnetospheres} \cite{mic91} and {\it The 
Physics of the Pulsar Magnetosphere} by Beskin, Gurevich \& Istomin 
\cite{bgi93}.  Our summary of evolutionary aspects serves merely as 
a primer to the vast body of literature available.  The reader is 
referred to the excellent review by Bhattacharya \& van~den Heuvel 
\cite{bv91} for further insights.
For an excellent overview of pulsar distance measurements and their
implications, see the review by Weisberg \cite{wei96}.

Pulsar resources available on the {\it Internet} are continually
becoming more extensive and useful. Good starting points for
pulsar-surfers are the pages maintained at Arecibo \cite{aopsr}, 
Berkeley \cite{bkypsr}, Bonn \cite{mpipsr}, Jodrell Bank \cite{jodpsr},
Princeton \cite{pripsr}, Swinburne \cite{swinpsr} and Sydney 
\cite{mbeampsr}.

\clearpage
\section{The Galactic Pulsar Population}
\label{sec:gal}

Soon after the discovery of pulsars, it was realised that the observed
sample is heavily biased towards the brighter objects that are the
easiest to detect. What we observe therefore most likely represents
only the tip of the iceberg of a much larger underlying population
\cite{go70}.  The extent to which the sample is incomplete is well
demonstrated by the projection of pulsars onto the Galactic plane and
their cumulative number distribution as a function of distance shown
in Fig.~\ref{fig:incomplete}. Although the clustering of sources
around the Sun seen in the left panel of Fig.~\ref{fig:incomplete}
would be consistent with Ptolemy's geocentric
picture of the heavens, it is clearly at variance with what we now
know about the Galaxy, where the massive stars show a radial
distribution about the Galactic center. 

\begin{figure}[hbt]
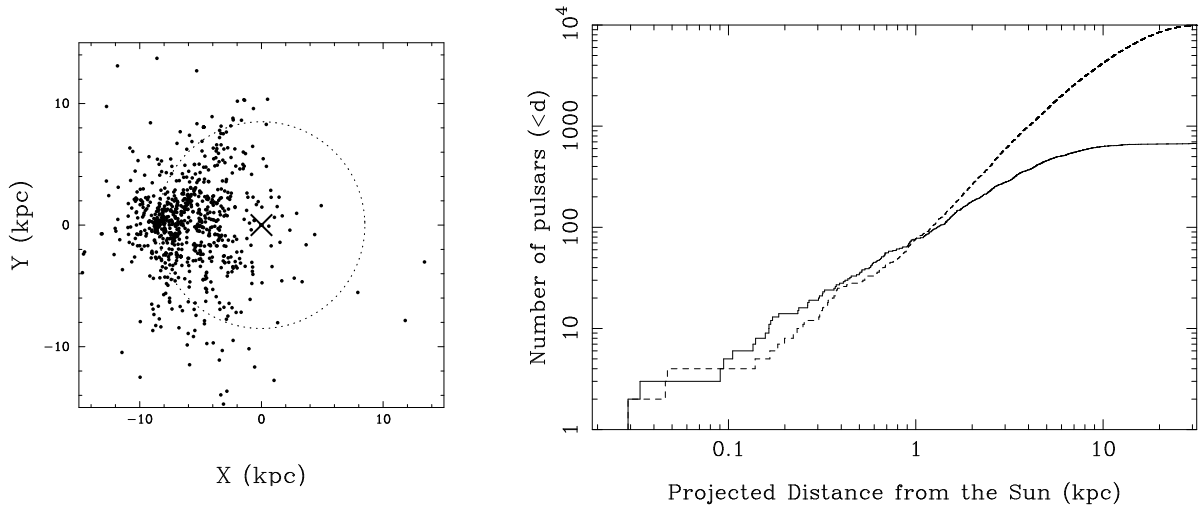

\setlength{\unitlength}{1in}
\begin{picture}(0,2.5)
\put(-0.1,+2.9){\includegraphics{xy.ps}}
\put(+2.3,+2.9){\includegraphics{cdist2.fig}}
\end{picture}
\caption[]
{\sl
Left: The sample of radio pulsars from the Princeton catalog
\cite{pripsr} projected onto the Galactic plane. The Galactic center 
is at (0,0) and the Sun is at (--8.5,0).  Right: Cumulative number 
of observed pulsars (solid line) as a function of projected distance, 
$d$. The dashed line shows the expected distribution for a model 
population (see text).
}
\label{fig:incomplete}
\end{figure}

The extent to which the pulsar sample is incomplete is shown in the
right panel of Fig.~\ref{fig:incomplete} where the cumulative number
of pulsars is plotted as a function of the projected distance from the
Sun. The observed distribution is compared to the
expected distribution for a simple model population in
which there are errors in the distance scale, but no selection
effects.  We see that the observed sample becomes strongly deficient
in terms of the number of sources for distances beyond a few kpc.
We now discuss the main selection effects that hamper the detection 
of pulsars in some detail.

\subsection{Selection effects in pulsar searches}
\label{sec:selfx}

\subsubsection{The inverse square law and survey thresholds}
\label{sec:invsq}

The most prominent selection effect at play in the observed pulsar
sample is the inverse square law, {\it i.e.}~for a given intrinsic
luminosity\footnote{Pulsar astronomers usually define the luminosity
$L = S d^2$, where $S$ is the mean flux density at 400 MHz (a standard
observing frequency) and $d$ is the distance derived from the DM 
(\S \ref{sec:dist}).},
the observed flux density falls off as the inverse square of the
distance.  This results in the observed sample being dominated by
nearby and/or bright objects.  Beyond distances of a few kpc from the
Sun, the apparent flux density falls below the flux thresholds $S_{\rm
min}$ of most surveys. Following \cite{dss+84}, we write:
\begin{equation}
\label{equ:defsmin}
S_{\rm min} = {\rm SNR}_{\rm min} \,
\left(\frac{T_{\rm rec}+T_{\rm sky}}{\rm K}\right) \, 
\left(\frac{G}{{\rm K \, Jy}^{-1}}\right)^{-1} \,
\left(\frac{\Delta \nu}{\rm MHz}\right)^{-1/2} \,  
\left(\frac{t_{\rm int}}{\rm s}\right)^{-1/2} \,
\left(\frac{W}{P-W}\right)^{1/2} \, {\rm mJy},
\end{equation}
where SNR$_{\rm min}$ is the threshold signal-to-noise
ratio, $T_{\rm rec}$ and $T_{\rm sky}$ are the receiver and sky noise
temperatures, $G$ is the gain of the antenna, $\Delta \nu$ is the
observing bandwidth, $t_{\rm int}$ is the integration time, $W$ is the
detected pulse width and $P$ is the pulse period. 

\subsubsection{Pulse dispersion and scattering}
\label{sec:dispandscatt}

It follows from Equation \ref{equ:defsmin} that the sensitivity decreases as
$W/(P-W)$ increases. Also note that if $W \gapp P$, the pulsed signal
is smeared into the background emission and is no longer detectable,
regardless of how luminous the source may be. The detected pulse
width $W$ will be broader than the intrinsic value largely as a result
of pulse dispersion and scattering by free electrons in the interstellar
medium. As discussed above, the dispersive smearing scales as $\Delta
\nu/\nu^3$, where $\nu$ is the observing frequency. This can largely
be removed by dividing the pass-band into a number of channels and
applying successively longer time delays to higher frequency channels
{\it before} summing over all channels to produce a sharp profile.
This process is known as incoherent dedispersion.

The smearing across the individual frequency channels, however, still
remains and becomes significant at high dispersions when searching for
short-period pulsars.  Multi-path scattering results in a one-sided
broadening due to the delay in arrival times which scales roughly as
$\nu^{-4}$, which can not be removed by instrumental means.  A simple
scattering model is shown in Fig. \ref{fig:scatt} in which the
scattering electrons are assumed to lie in a thin screen between the
pulsar and the observer \cite{sch68}.

\begin{figure}[hbt]
\setlength{\unitlength}{1in}
\begin{picture}(0,2)
\put(1.8,0){\includegraphics{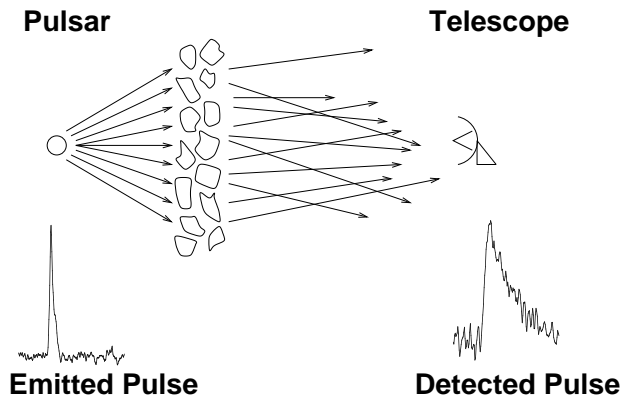}}
\end{picture}
\caption[]
{\sl
Pulse scattering caused by irregularities in the interstellar medium.
The difference in path lengths and therefore in arrival times of the
scattered rays results in a ``scattering tail'' in the observed pulse
profile which lowers its signal-to-noise ratio.
}
\label{fig:scatt}
\end{figure}

Dispersion and scattering are most severe for distant pulsars in
the inner Galaxy where the number of free electrons along the line of
sight becomes large. The strong frequency dependence of both effects
means that they are considerably less of a problem for surveys at
observing frequencies $\gapp 1400$ MHz \cite{clj+92,jlm+92} compared 
to the usual 400 MHz search frequency.  An added bonus for such
observations is the reduction in $T_{\rm sky}$, since the spectral
index of the non-thermal Galactic emission is about --2.8
\cite{lmop87}. Pulsars themselves have steep radio spectra. Typical
spectral indices are --1.6 \cite{lylg95}, so that flux
densities are roughly an order of magnitude lower at 1400 MHz compared to 400
MHz. Fortunately, this can usually be compensated for by the 
use of larger receiver bandwidths at higher radio frequencies. For
example, the 1370-MHz system at Parkes has a bandwidth of 288 MHz \cite{lcm+00}
compared to the 430-MHz system, where 32 MHz is available \cite{mld+96}.

\subsubsection{Orbital acceleration}
\label{sec:accn}

Standard pulsar searches use Fourier techniques to search for {\it
a-priori} unknown periodic signals and usually assume that the
apparent pulse period remains constant throughout the observation. For
searches with integration times much greater than a few minutes this
assumption is only valid for solitary pulsars, or those in binary
systems where the orbital periods are longer than about a day. For
shorter-period binary systems, as noted by Johnston \& Kulkarni
\cite{jk91}, the Doppler-shifting of the period results in a spreading
of the signal power over a number of frequency bins in the Fourier
domain, leading to a reduction in signal-to-noise ratio.  An observer
will perceive the frequency of a pulsar to shift by an amount $aT/(Pc)$,
where $a$ is the (assumed constant) line-of-sight acceleration during
the observation of length $T$, $P$ is the (constant) pulse period in
its rest frame and $c$ is the speed of light. Given that the width of
a frequency bin is $1/T$, we see that the signal will drift into more
than one spectral bin if $aT^2/(Pc)>1$. Survey sensitivities to
rapidly-spinning pulsars in tight orbits are therefore significantly
compromised when the integration times are large.

\begin{figure}[hbt]
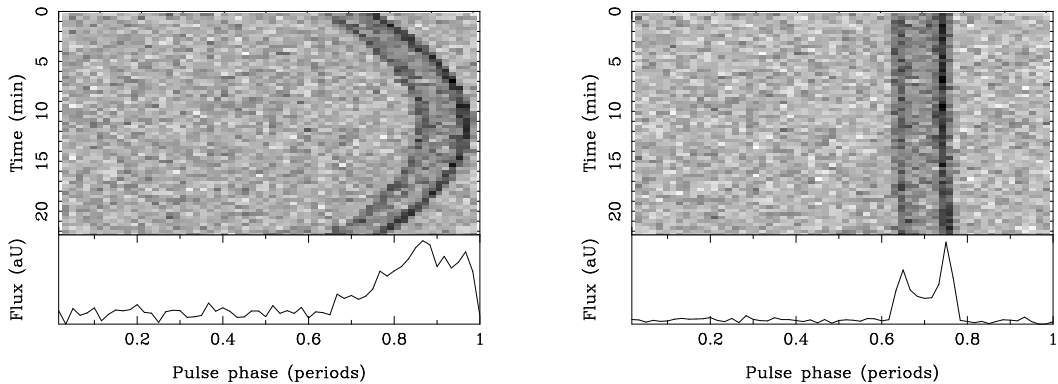

\setlength{\unitlength}{1in}
\begin{picture}(0,2)
\put(0.0,2.5){\includegraphics{1913_raw.ps}}
\put(3.0,2.5){\includegraphics{1913_cor.ps}}
\end{picture}
\caption[] 
{\sl
Left: a 22.5-min Arecibo observation of the binary pulsar B1913+16.
The assumption that the pulsar has a constant period during this time
is clearly inappropriate given the drifting in phase of the pulse
during the integration (grey scale plot). Right: the same observation
after applying an acceleration search. This shows the effective
recovery of the pulse shape and a significant improvement 
in the signal-to-noise ratio.
}
\label{fig:1913acc}
\end{figure}

As an example of this effect, as seen in the time domain,
Fig.~\ref{fig:1913acc} shows a 22.5-min search mode observation of Hulse
\& Taylor's famous binary pulsar B1913+16 \cite{ht75a,tw82,tw89}.
Although this observation covers only about 5\% of the orbit (7.75
hr), the effects of the Doppler smearing on the pulse signal are very
apparent. While the standard search code (seeking constant periodicity)
nominally detects the pulsar with a
signal-to-noise ratio of 9.5 for this observation, it is clear that
the Doppler shifting of the pulse period seen in the individual
sub-integrations results in a significant reduction in signal-to-noise.

It is clearly desirable to employ a technique to recover the loss in
sensitivity due to Doppler smearing. One such technique, the so-called
``acceleration search'' \cite{mk84}, assumes the pulsar has a constant
acceleration during the integration.  Each time series can then be
re-sampled to refer it to the frame of an inertial observer using the
Doppler formula to relate a time interval, $\tau$, in the pulsar frame
to that in the observed frame at time $t$, as $\tau(t) \propto ( 1 +
at/c )$. Searching over a range of accelerations is desirable to find
the time series for which the trial acceleration most closely matches
the true value. In the ideal case, a time series is produced with a
signal of constant period for which full sensitivity is recovered (see
right panel of Fig.~\ref{fig:1913acc}). Anderson et al.~\cite{agk+90} 
used this technique to find PSR B2127+11C, a double neutron star
binary in M15 which has parameters similar to B1913+16. Camilo et
al.~\cite{clf+00} have recently applied the same technique to
47~Tucanae to discover 9 binary pulsars, including one in a 96-min
orbit around a low-mass (0.15 M$_{\odot}$) companion. This is
currently the shortest binary period for any known radio pulsar.

For the shortest orbital periods, the assumption of a constant
acceleration during the observation clearly breaks down. Ransom et
al.~\cite{rgh+01} have developed a particularly efficient algorithm
for finding binaries whose orbits are so short that many orbits can
take place during an integration. This phase modulation technique
exploits the fact that the pulses are modulated by the orbit to create
a family of periodic sidebands around the nominal spin period of the
pulsar. This technique has already been used to discover a 1.7-hr
binary pulsar in NGC 6544~\cite{rgh+01}. The existence of these
short-period radio pulsar binaries, as well as the 11-min X-ray binary
X1820$-$303 in NGC~6624 \cite{spw87}, implies that there must be many
more short-period binaries containing radio or X-ray pulsars in globular 
clusters that are waiting to be discovered by more sensitive searches.

\subsection{Correcting the observed pulsar sample}
\label{sec:corsamp}

\subsubsection{Scale Factor determination}
\label{sec:sfacts}

Now that we have a flavour for the variety and severity of the
selection effects that plague the observed sample of pulsars, how
do we decouple these effects to form a less biased picture of the true
population of objects? A very useful technique, first employed by
Phinney \& Blandford and Vivekanand \& Narayan \cite{pb81,vn81}, is to
define a scaling factor $\xi$ as the ratio of the total Galactic
volume weighted by pulsar density to the volume in which a pulsar
could be detected by the surveys:
\begin{equation}
\label{equ:sfac}
\xi(P,L) = \frac{\int \int_{\rm Galaxy} \, \Sigma(R,z) \, R \, \, dR \, dz}
           {\int \int_{P,L} \, \Sigma(R,z) \, R \, \, dR \, dz}.
\end{equation}
In this expression, $\Sigma(R,z)$ is the assumed pulsar distribution
in terms of galactocentric radius $R$ and height above the Galactic
plane $z$. Note that $\xi$ is primarily a function of period $P$ and
luminosity $L$ such that short period/low-luminosity pulsars have
smaller detectable volumes and therefore higher $\xi$ values than
their long period/high-luminosity counterparts.  This approach is
similar to the classic $V/V_{\rm max}$ technique first used to correct
observationally-biased samples of quasars \cite{sch68b}.

This technique can be used to estimate the total number of
active pulsars in the Galaxy.  In practice, this is achieved by
calculating $\xi$ for each pulsar separately using a Monte Carlo
simulation to model the volume of the Galaxy probed by the major
surveys \cite{nar87}. For a sample of $N_{\rm obs}$ observed pulsars
above a minimum luminosity $L_{\rm min}$, the total number of
pulsars in the Galaxy with luminosities above this value is simply
\begin{equation}
\label{equ:ngal}
\sum_{i=1}^{N_{\rm obs}} \frac{\xi_i}{f_i},
\end{equation}
where $f$ is the model-dependent ``beaming fraction'' discussed below in
\S \ref{sec:beaming}. Monte Carlo simulations of the pulsar population
incorporating the aforementioned selection effects have shown this
method to be reliable, as long as $N_{\rm obs}$ is reasonably large
\cite{lbdh93}.

\subsubsection{The small-number bias}
\label{sec:smallnumber}

For small samples of observationally-selected objects, the detected
sources are likely to be those with larger-than-average luminosities.
The sum of the scale factors (Equation \ref{equ:ngal}), therefore,
will tend to underestimate the true size of the population. This
``small-number bias'' was first pointed out by Kalogera et
al.~\cite{kal00,knst01} for the sample of double neutron star binaries where
we know of only three clear-cut examples (\S \ref{sec:nsns}).  Only when the
number of sources in the sample gets past 10 or so does the sum of the
scale factors become a good indicator of the true population size.

\begin{figure}[htb]
\setlength{\unitlength}{1in}
\begin{picture}(0,2.5)
\put(0.1,-0.1){\includegraphics{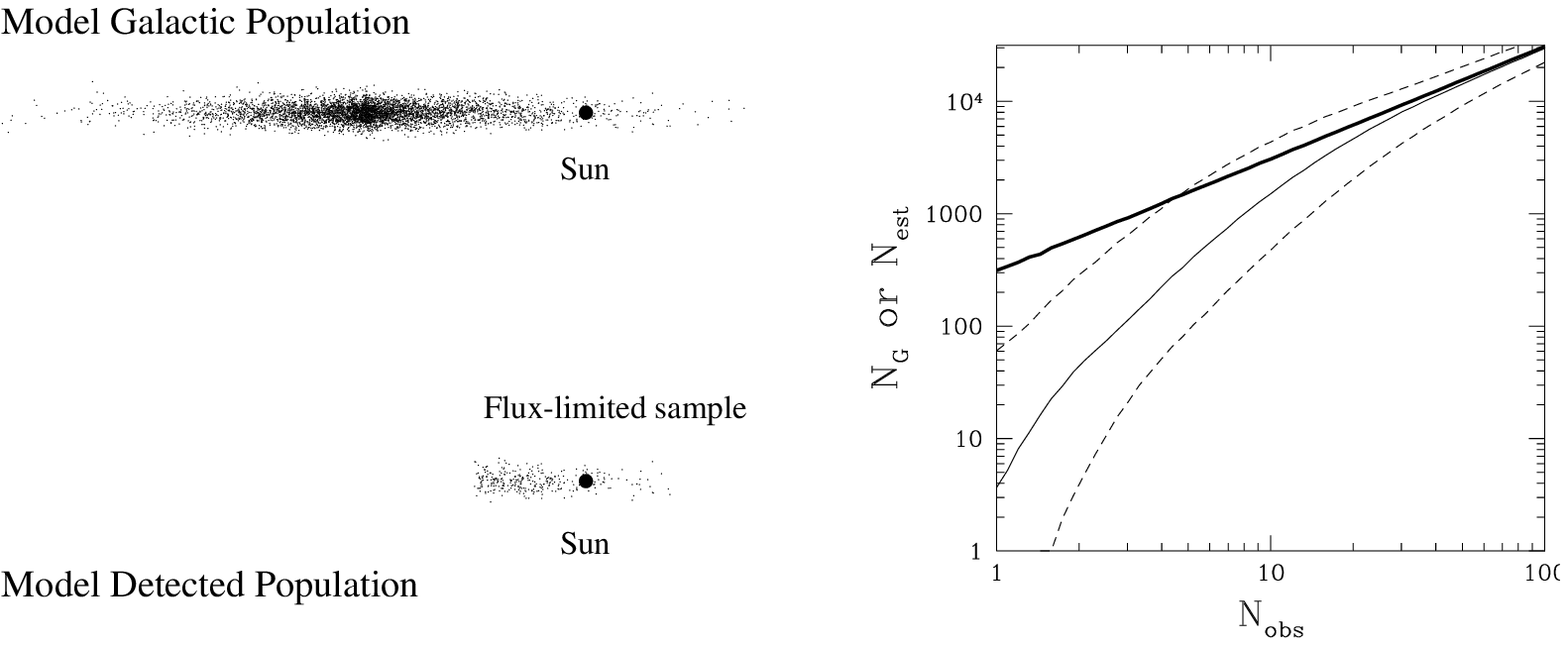}}
\end{picture}
\caption{\sl
Small-number bias of the scale factor estimates derived from a
synthetic population of sources where the true number of sources is
known. Left: an edge-on view of a model Galactic source population.
Right: the thick line shows $N_{\rm G}$, the true number of objects in
the model Galaxy, plotted against $N_{\rm obs}$, the number detected by
a flux-limited survey. The thin solid line shows $N_{\rm est}$, the
median sum of the scale factors, as a function of
$N_{\rm obs}$ from a large number of Monte-Carlo trials.  Dashed lines
show 25 and 75\% percentiles of the $N_{\rm est}$ distribution.
}
\label{fig:smallnumber}
\end{figure}

\subsubsection{The beaming correction}
\label{sec:beaming}

The ``beaming fraction'' $f$ in Equation \ref{equ:ngal}
is simply the fraction of $4\pi$ steradians swept
out by the radio beam during one rotation. Thus $f$ gives the probability
that the beam cuts the line-of-sight of an arbitrarily positioned
observer.  A na\"{\i}ve estimate of $f$ is 20\%; this assumes a beam
width of $\sim 10^{\circ}$ and a randomly distributed inclination
angle between the spin and magnetic axes \cite{tm77}. Observational
evidence suggests that shorter period pulsars have wider beams and
therefore larger beaming fractions than
their long-period counterparts \cite{nv83,lm88,big90b,tm98}.
It must be said, however, that a consensus on the beaming
fraction-period relation has yet to be reached. 
This is shown in Fig.~\ref{fig:bfracts} where we compare the period 
dependence of $f$ as given by a number of models.
Adopting the Lyne \& Manchester model, pulsars with periods $\sim 100$
ms beam to about 30\% of the sky compared to the Narayan \& Vivekanand
model in which pulsars with periods below 100 ms beam to the entire sky. 

\begin{figure}[hbt]
\setlength{\unitlength}{1in}
\begin{picture}(0,2.8)
\put(0.8,3.45){\includegraphics{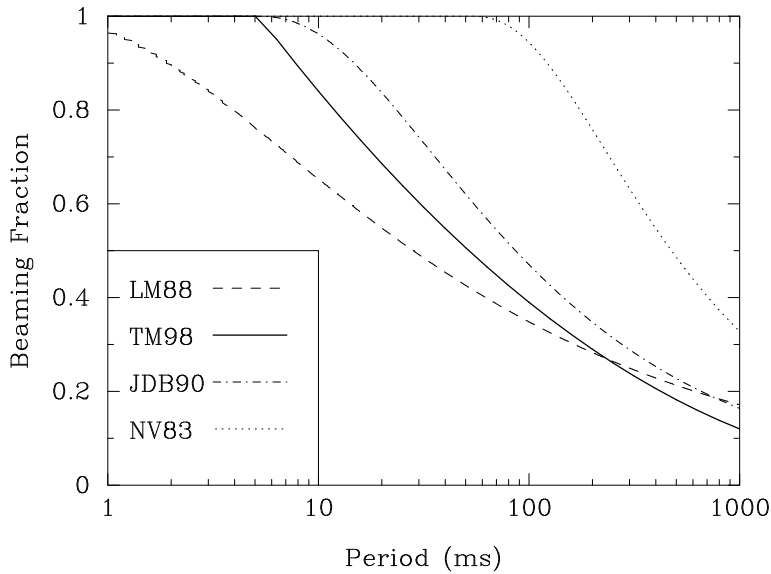}}
\end{picture}
\caption[]
{\sl
Beaming fraction plotted against pulse period for four
different beaming models: Tauris \& Manchester 1998 (TM88; \cite{tm98}),
Lyne \& Manchester 1988 (LM88; \cite{lm88}), Biggs 1990
(JDB90; \cite{big90b}) and Narayan \& Vivekanand 1983 (NV83 \cite{nv83}).
}
\label{fig:bfracts}
\end{figure}

When most of these beaming models were originally proposed, the sample of
millisecond pulsars was $\lapp$ 5 and hence their predictions about
the beaming fractions of short-period pulsars relied largely on
extrapolations from the normal pulsars.  A recent analysis of a large
sample of millisecond pulsar profiles by Kramer et al.~\cite{kxl+98} 
suggests that the beaming fraction of millisecond pulsars lies between 
50 and 100\%. 

\subsection{The population of normal and millisecond pulsars}
\label{sec:nmsppop}

\subsubsection{Luminosity distributions and local number estimates}
\label{sec:lumfuns}

The most recent use of the scale factor approach to derive the
characteristics of the true normal and millisecond pulsar populations
is based on the sample of pulsars within 1.5 kpc of the Sun
\cite{lml+98}. The rationale for this cut-off is that, within this
region, the selection effects are well understood and easier to
quantify by comparison with the rest of the Galaxy. These calculations
should give reliable estimates for the {\it local pulsar population}.

\begin{figure}[hbt]
\setlength{\unitlength}{1in}
\begin{picture}(0,2)
\put(-0.05,2.2){\includegraphics{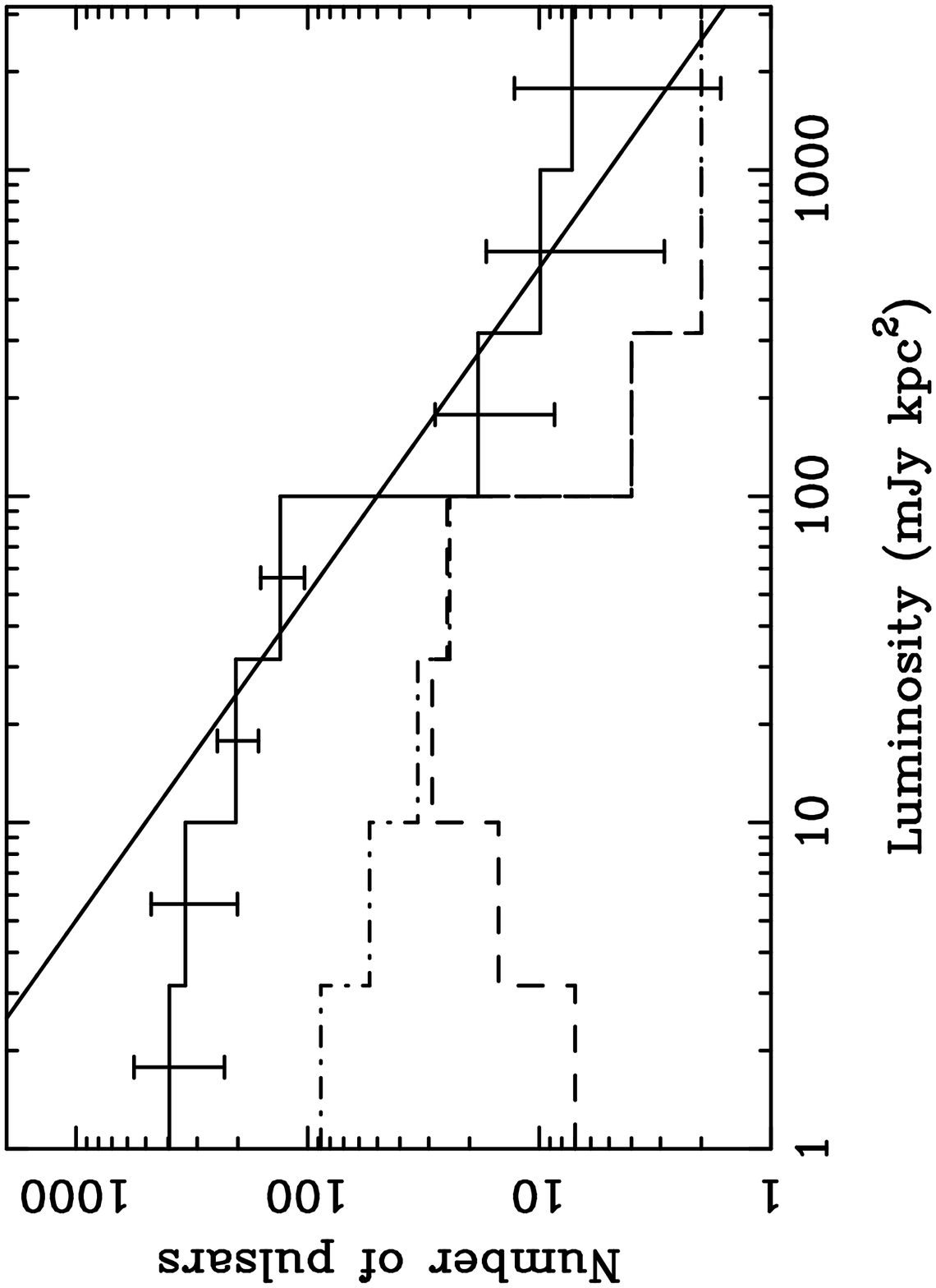}}
\put(+2.90,2.2){\includegraphics{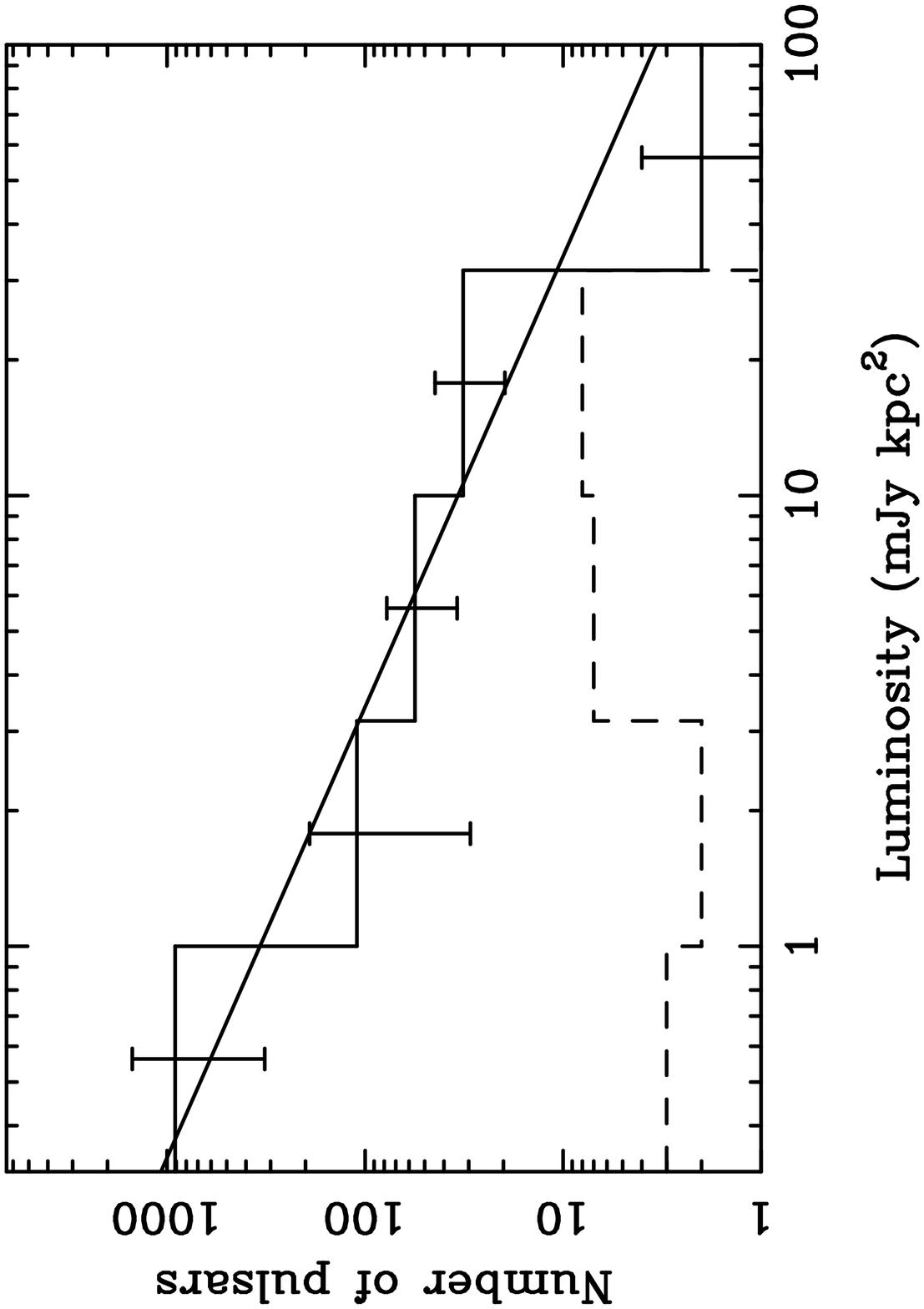}}
\end{picture}
\caption[]
{\sl
Left: The corrected luminosity distribution (solid histogram with
error bars) for normal pulsars.  The corrected distribution {\sl
before} the beaming model has been applied is shown by the dot-dashed
line.  Right: The corresponding distribution for millisecond
pulsars. In both cases, the observed distribution is shown by the
dashed line and the thick solid line is a power law with a slope of
--1.  The difference between the observed and corrected distributions
highlights the severe under-sampling of low-luminosity pulsars.
}
\label{fig:lumfuns}
\end{figure}

The luminosity distributions obtained from this analysis are shown in
Fig.~\ref{fig:lumfuns}. For the normal pulsars, integrating the
corrected distribution above 1 mJy kpc$^2$ and dividing by $\pi \times
(1.5)^2$ kpc$^2$ yields a local surface density, assuming Biggs'
beaming model \cite{big90b}, of $156 \pm 31$ pulsars kpc$^{-2}$. The
same analysis for the millisecond pulsars, assuming a mean beaming
fraction of 75\% \cite{kxl+98}, leads to a local surface
density of $38 \pm 16$ pulsars kpc$^{-2}$ for luminosities above
1 mJy kpc$^2$.

\subsubsection{Galactic population and birth-rates}
\label{sec:psrpop}

Integrating the local surface densities of pulsars over the whole
Galaxy requires a knowledge of the presently rather uncertain
Galactocentric radial distribution \cite{lmt85,joh94}. One approach is
to assume that pulsars have a radial distribution similar to that of
other stellar populations and scale the local number density with 
this distribution to estimate the total Galactic population. 
The corresponding local-to-Galactic scaling is
1000$\pm$250 kpc$^2$ \cite{rv89}. With this approach we estimate there
to be $\sim 160,000$ active normal pulsars and $\sim 40,000$
millisecond pulsars in the Galaxy.  Based on these estimates, we are
in a position to deduce the corresponding rate of formation or
birth-rate.  From the $P$--$\dot{P}$ diagram in
Fig.~\ref{fig:ppdot}, we infer a typical lifetime for normal pulsars
of $\sim 10^{7}$ yr, corresponding to a Galactic birth rate of 
$\sim 1$ per 60 yr --- consistent with the rate of supernovae
\cite{vt91}. As noted in \S \ref{sec:nms}, the
millisecond pulsars are much older, with ages close to that of
the Universe, $\tau_u$ (we assume here $\tau_u=10$ Gyr \cite{jfk98}).
Taking the maximum age of the millisecond pulsars to be $\tau_u$, we
infer a mean birth rate of at least $4 \times 10^{-6}$ yr$^{-1}$. This
is consistent, within the uncertainties, with the birth-rate of
low-mass X-ray binaries \cite{lnl+95}.

\subsubsection{Implications for gravitational wave detectors}
\label{sec:psrgws}

The estimates of the local surface density of active pulsars allow us
to deduce the likely distance to the nearest neutron star to Earth.
For the combined millisecond and normal pulsar populations, with a
surface density of $193 \pm 35$ pulsars kpc$^{-2}$, the nearest
neutron star is thus likely to be $\lapp 40$ pc.  This number is of
interest to those building gravitational wave detectors, since it
determines the likely amplitude of gravitational waves emitted from
nearby rotating neutron stars \cite{sch00}. According to Thorne \cite{tho96},
currently planned detectors will be able to detect neutron stars with
ellipticities greater than $7.5 \times 10^{-11} (Pd)^2$,
where $P$ is the rotation period in ms and $d$ is the distance in kpc.
The recent probable detection of free precession in the radio pulsar B1828$-$11
\cite{sls01} does indicate that ellipticities exist in neutron stars
so that nearby objects may be continuous sources of gravitational radiation.

Thus, in order to detect small ellipticities, nearby sources with
short spin periods are required. One of the best candidates is the
nearby 5.75-ms pulsar J0437-4715 \cite{jlh+93}. At a distance of $178
\pm 26$ pc \cite{sbm+97} this is currently the closest known millisecond 
pulsar to the Earth. The closest known neutron star is RX J185635$-$3754
discovered in the ROSAT all-sky survey \cite{wfn96}. Multi-epoch
HST observations show that this isolated neutron star is located at a
distance of $61\pm9$ pc \cite{wal00}. In keeping with other radio-quiet 
isolated neutron stars, the period of this pulsar is likely to be 
several seconds \cite{nt99}.

\subsection{The population of relativistic binaries}
\label{sec:relpop}

Of particular interest to the astronomical community at large are
the numbers of relativistic binary systems in the Galaxy. Systems
involving neutron stars are: double neutron star binaries,
neutron star--white dwarf binaries and neutron star--black hole
binaries. Interest in these systems is two-fold: (1) to test the 
predictions of general relativity against alternative theories
of strong-field gravity using the radio pulsar as a highly
stable clock moving in the strong gravitational field; (2) to
detect strong gravitational wave emission from coalescing
binaries with upcoming gravitational-wave observatories like 
GEO600 and LIGO, VIRGO and TAMA\cite{tho96,sch98}.

Although no radio pulsar in orbit around a black hole companion has so
far been observed, we now know of several double neutron star and
neutron--star white dwarf binaries which will merge due to
gravitational wave emission within a reasonable time-scale.  The
merging time $\tau_g$ of a binary system containing two compact objects due to
the emission of gravitational radiation can be calculated
from the following formula which requires only the component
masses and current orbital period $P_b$ and eccentricity $e$:
\begin{equation}
\label{equ:tgw}
\tau_g \simeq 10^7 \, {\rm yr} \, \,
\left(\frac{P_b}{\rm hr}\right)^{8/3} 
\left(\frac{m_1 + m_2}{\rm M_{\odot}}\right)^{1/3} 
\left(\frac{\mu}{\rm M_{\odot}}\right)^{-1} 
(1-e^2)^{7/2}.
\end{equation}
Here $m_{1,2}$ are the masses of the two stars and
$\mu=m_1m_2/(m_1+m_2)$.  This formula is a good analytic approximation
(within a few percent) to the numerical solution of the exact
equations for $\tau_g$ in the original papers by Peters \& Mathews
\cite{pet64,pm63}.  In the following subsections we review current
knowledge on the population sizes and merging rates of such binaries
where one component is visible as a radio pulsar.

\subsubsection{Double neutron star binaries}
\label{sec:nsns}

As noted in \S \ref{sec:nms}, double neutron star (DNS) binaries
are expected to be rare since the binary system has to survive
two supernova explosions. This expectation is certainly borne out by
radio pulsar searches which have revealed only
three certain DNS binaries so far: 
PSRs B1534+12 \cite{wol91a}, B1913+16
\cite{ht75a} and B2127+11C \cite{pakw91}. Although we cannot see the companion
neutron star in any of these systems, we are ``certain'' of the
identification from the precise measurements of the component masses
via relativistic effects measured in pulsar timing observations (see
\S \ref{sec:timobs}). The spin and orbital parameters of these
pulsars are listed in Table \ref{tab:nsns}.
\begin{table}[hbt]
\footnotesize
\begin{center}
\begin{tabular}{lrrrrrr}
\hline
\hline
&J1518+4904&B1534+12&J1811$-$1736&B1820$-$11&B1913+16&B2127+11C\\
\hline
$P$ (ms)                 &40.9      &37.9    &104.2 &279.8 &59.0    &30.5\\
$P_b$ (d)                &8.6       &0.4     &18.8  &357.8 &0.3     &0.3\\
$e$                      &0.25      &0.27    &0.83  &0.79  &0.62    &0.68\\
$\tau_c$  ($10^8$ y)     &200       &2.5     &970   &0.04  &1.1     &0.97\\
$\tau_g$ ($10^8$ y)      &$\gg\tau_u$&27&$\gg\tau_u$&$\gg\tau_u$&3.0&2.2 \\
Masses measured?         &No        &Yes     &No    &No    &Yes     &Yes \\
\hline
\end{tabular}
\end{center}
\caption[]
{\sl
Known DNS binaries and candidates.  Listed are the pulse period $P$,
the orbital period $P_b$, the orbital eccentricity $e$, the
pulsar characteristic age $\tau_c$, the expected binary coalescence
time-scale $\tau_g$ due to gravitational wave emission calculated from
Equation \ref{equ:tgw}. Cases for which $\tau_g$ is a factor of 100 or
more greater than the age of the Universe are listed as $\gg\tau_u$. 
To distinguish between definite and candidate DNS systems, we also 
list whether the masses of both components have been determined.
}
\label{tab:nsns}
\end{table}
Also listed in Table \ref{tab:nsns} are three further DNS
candidates with eccentric orbits and large mass functions
but for which there is presently not sufficient component
mass information to confirm their nature. 

Despite the uncertainties in identifying DNS binaries, for the
purposes of determining the Galactic merger rate,
the systems for which $\tau_g$ is less than $\tau_u$ (i.e.~PSRs
B1534+12, B1913+16 and B2127+11C) are primarily of interest. Of these
PSR B2127+11C is in the process of being ejected from the globular
cluster M15 \cite{pakw91,ps91} and is thought to make only a negligible
contribution to the merger rate \cite{phi91}.  The general approach with
the remaining two systems is to derive scale factors for each object
(as outlined in \S \ref{sec:sfacts})
and then divide these by a reasonable estimate for the lifetime. In
what follows we summarize the main studies of this kind. The most
comprehensive investigation of the DNS binary population to date is
the recent study by Kalogera et al.~(hereafter KNST; \cite{knst01}).

As discussed in \S \ref{sec:sfacts}, scale factors are dependent on
the assumed pulsar distribution. The key parameter here is the scale
height of the population with respect to the Galactic plane which
itself is a function of the velocity distribution of the population. 
KNST examined this dependence in detail and found scale heights in the
range 0.8--1.7 kpc. Based on this range, KNST revised earlier scale 
factor estimates \cite{cl95} to 145--200 for B1534+12 and 45--60 for
PSR B1913+16. As mentioned in \S \ref{sec:smallnumber} scale factors 
calculated from a small sample of objects are subject to a significant
bias. KNST find the bias in their sample to be anywhere between 2 and
200. This boosts the scale factors to the range 190--40000 for
B1534+12 and 90--12000 for B1913+16.

The above scale factors also require a beaming correction. As noted
in \S \ref{sec:beaming}, current radio pulsar beaming models
vary considerably. Fortunately, for the two pulsars under consideration,
detailed studies of the beam sizes \cite{aptw96,kra98,wt00} lead KNST
to conclude that both pulsars beam to only about a sixth of the entire
sky. The beaming-corrected numbers suggest a total of between 1680 and
312,000 active DNS binaries in our Galaxy. Many of these systems will
be extremely faint objects.  These estimates are dominated by
the small-number bias factor.  KNST's study highlights the importance 
of this effect.

Some debate exists about what is the most reasonable estimate of the
lifetime. Phinney \cite{phi91} defines this as the sum of the pulsar's
spin-down age plus $\tau_g$ defined above. A few years later,
van den Heuvel and myself argued \cite{vl96} that a more likely
estimate can be obtained by appealing to
steady-state arguments where we expect sources to be created at the
same rate at which they are merging. The mean lifetime was then found
to be about three times the current spin-down age.  This argument
does, however, depend on the luminosity evolution of radio pulsars
which is currently only poorly understood.  Arzoumanian, Cordes \&
Wasserman \cite{acw99} used kinematic data to constrain the most likely
ages of the DNS binaries. They note that the remaining detectable
lifetime should also take account of the reduced detectability at
later epochs due to acceleration smearing as the DNS binary becomes
more compact due to gravitational wave emission. KNST concluded that
the lifetimes are dominated by the latter time-scale which, following
Arzoumanian et al., they took to be the time for the orbital period to
halve.  The resulting lifetimes are $2.5\times10^{9}$ yr for B1534+12
and $2.5\times10^{8}$ yr for B1913+16.

Taking these number and lifetime estimates, KNST find the Galactic
merger rate of DNS binaries to range between $3\times10^{-6}$ and
$4\times10^{-4}$ yr$^{-1}$. Extrapolating this number out to include
DNS binaries detectable by LIGO in other galaxies \'{a} la Phinney 
\cite{phi91} KNST find the expected event rate to be $<0.25$ yr$^{-1}$
for LIGO--I and 2--1300 yr$^{-1}$ for LIGO--II.  Thus, despite the
uncertainties, it seems that the prospects for detecting gravitational-wave 
emission from DNS inspirals in the near future are most promising.  

\subsubsection{White dwarf--neutron star binaries}
\label{sec:nswd}

The population of white dwarf--neutron star (WDNS) sources containing
a young radio pulsar has only
recently been confirmed by observers following the identification of
a white dwarf companion to the binary pulsar B2303+46 by van Kerkwijk
et al.~\cite{vk99}.  Previously, this eccentric binary pulsar was
thought to be an example of a DNS binary in which the visible pulsar
is the second-born neutron star \cite{std85,lb90}.  The optical
identification rules this out and now strongly suggests a scenario in
which the white dwarf was formed first.  In this case, material was
transfered onto the secondary during the giant phase of the primary
so that the secondary became massive enough to form a neutron
star \cite{vk99,ts00}.

PSR B2303+46 has a long orbital period and does not contribute
significantly to the overall merger rate of WDNS binaries.  The new
discovery of PSR J1141--6545 \cite{klm+00}, which will merge due to
gravitational-wave emission within 1.3 Gyr, is suggestive of a large
population of similar binaries.  This is particularly compelling when
one considers that the radio lifetime of the visible pulsar is only a
fraction of total lifetime of the binary before coalescence due to
gravitational-wave emission.  Edwards \& Bailes \cite{eb01} estimate
there to be 850 WDNS binaries within 3 kpc of the Sun which will merge
within $\tau_u$.

Population syntheses by Tauris \& Sennels \cite{ts00} suggest that the
formation rate of WDNS binaries is between 10--20 times that of DNS
binaries. Based on the merging rate estimates for DNS binaries
discussed in the previous section, this translates to a merging rate
of WDNS binaries of between $3\times10^{-5}$ yr$^{-1}$ and
$8\times10^{-3}$ yr$^{-1}$.  In summary, although statistics are
necessarily poor at this stage, coalescing WDNS binaries look to be
very promising sources for gravitational wave detectors.

\subsection{Going further}

Studies of pulsar population statistics represent a large proportion of 
the pulsar literature. During this section we have tried to cite many of the
key papers in this field. Good starting points for further reading can be 
found in other review articles \cite{mic91,bv91,ls98}. Our coverage of
compact object coalescence rates has concentrated on empirical methods
and we have hopefully convinced the reader that these are fair
and straightforward. 

An alternative approach is to undertake a full-blown Monte Carlo
simulation of the most likely evolutionary scenarios described in \S
\ref{sec:evolution}. In this ``scenario-machine'' approach, a
population of primordial binaries is synthesized with a number of
underlying distribution functions: primary mass, binary mass ratio,
orbital period distribution etc. The evolution of both stars is then
followed to give a predicted sample of binary systems of all the
various types. Since the full range of binary parameters is known, the
merger rates of each type of binary are then automatically predicted
by this model without the need to debate what the likely coalescence
times will be. Selection effects are not normally taken into account
in this approach. The final census is usually normalized to the
star formation rate.
Numerous examples of the scenario-machine approach (most often to
populations of binaries where one or both members are NSs) can be
found in the literature \cite{dc87,rom92,ty93}.  These include the
widely-cited code, developed by Lipunov and collaborators to perform
population syntheses of binary stars \cite{scenario}.
Although extremely instructive, the uncertain assumptions about
initial conditions, the physics of mass transfer and the kicks applied
to the compact object at birth result in a wide range of predicted
event rates which are currently broader than the empirical methods
\cite{kal00}. Ultimately, the detection
statistics from the gravitational wave detectors could provide far
tighter constraints on the DNS merging rate than the pulsar surveys
from which these predictions are made.

An excellent overview of gravitational-wave astronomy and the
detection of gravitational waves from inspiraling binaries is presented
by Thorne in his presentation at the centennial meeting of the 
American Physical Society which is available on-line \cite{thornetalk}.

\clearpage
\section{Pulsar Timing}
\label{sec:pultim}

It became clear soon after their discovery that pulsars are excellent
celestial clocks. In the original discovery paper \cite{hbp+68}, the
period of the first pulsar to be discovered, PSR B1919+21, was found
to be stable to one part in $10^7$ over a time-scale of a few
months. Following the discovery of the millisecond pulsar B1937+21 in
1982 \cite{bkh+82} it was demonstrated that its period could be
measured to one part in $10^{13}$ or better \cite{dtwb85}. This
unrivaled stability leads to a host of applications including time
keeping, probes of relativistic gravity and natural gravitational
wave detectors. 

\subsection{Observing basics}
\label{sec:timobs}

As each new pulsar is discovered, the standard practice is to add it
to a list of pulsars which are regularly observed at least once or
twice per month by large radio telescopes throughout the world.  The
schematic diagram shown in Fig.~\ref{fig:timing} summarises the
essential steps involved in such a ``time-of-arrival'' (TOA)
measurement. Incoming pulses emitted by the rotating neutron star
traverse the interstellar medium before being received by the radio
telescope. After amplification by high sensitivity receivers, the
pulses are de-dispersed and added to form a mean pulse profile.

\begin{figure}[hbt]
\setlength{\unitlength}{1in}
\begin{picture}(0,1.75)
\put(-0.7,3.5){\includegraphics{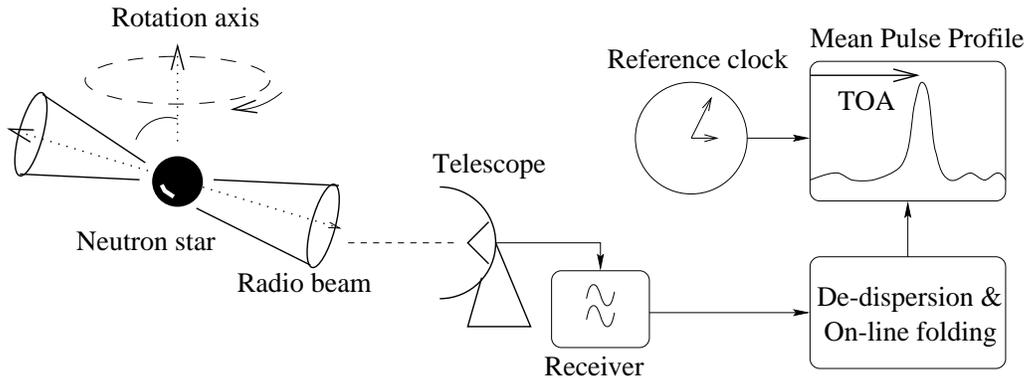}}
\end{picture}
\caption[]
{\sl
Schematic showing the main stages involved in pulsar timing observations.
}
\label{fig:timing}
\end{figure}

During the observation, the data regularly receive a time stamp,
usually based on a caesium time standard or hydrogen maser at 
the observatory plus a signal from the Global Positioning 
System of satellites (GPS; \cite{gps}). The TOA of this mean
pulse is then defined as the arrival time of some fiducal point on the
profile. Since the mean profile has a stable form at any given
observing frequency (\S \ref{sec:profs}), the TOA can be accurately
determined by a simple cross-correlation of the observed profile with
a high signal-to-noise ``template'' profile obtained from the
addition of many observations of the pulse profile at the particular
observing frequency.

Success in pulsar timing
hinges on how precisely the fiducial point can be determined. This
is largely dependent on the signal-to-noise ratio (SNR) of the mean
pulse profile. The uncertainty in a TOA measurement $\epsilon_{\rm TOA}$
is given roughly by the pulse width divided by the SNR. Using Equation
\ref{equ:defsmin}, we can express this as a fraction of the pulse period:
\begin{equation}
\label{equ:defsnr}
\frac{\epsilon_{\rm TOA}}{P} \simeq 
\left(\frac{T_{\rm rec}+T_{\rm sky}}{\rm K}\right) \, 
\left(\frac{G}{{\rm K \, Jy}^{-1}}\right)^{-1} \,
\left(\frac{\Delta \nu}{\rm MHz}\right)^{-1/2} \,  
\left(\frac{t_{\rm int}}{\rm s}\right)^{-1/2} \,
\left(\frac{W}{P}\right)^{3/2}.
\end{equation}
In this expression $T_{\rm rec}$ and $T_{\rm sky}$ are the receiver
and sky noise temperatures, $G$ is the gain of the antenna, $\Delta
\nu$ is the observing bandwidth, $t_{\rm int}$ is the integration
time, $W$ is the pulse width and $P$ is the pulse period (we assume $W
\ll P$).  Optimum results are thus obtained for observations of short
period pulsars with large flux densities and narrow duty cycles
($W/P$) using large telescopes with low-noise receivers and
large observing bandwidths.

One of the main problems of employing large bandwidths is pulse
dispersion. As discussed in \S \ref{sec:dist}, the velocity of the
pulsed radiation through the ionised interstellar medium is
frequency-dependent: pulses emitted at higher radio frequencies travel
faster and arrive earlier than those emitted at lower
frequencies. This process has the effect of ``stretching'' the pulse
across a finite receiver bandwidth, reducing the apparent
signal-to-noise ratio and therefore increasing $\epsilon_{\rm TOA}$.
For most normal pulsars, this process can largely be compensated for
by the incoherent de-dispersion process outlined in \S \ref{sec:selfx}.

To exploit the precision offered by millisecond pulsars, a more
precise method of dispersion removal is required. Technical
difficulties in building devices with very narrow channel bandwidths
require another dispersion removal technique. In the process of
coherent de-dispersion \cite{han71} the incoming signals are
de-dispersed over the whole bandwidth using a filter which has the
inverse transfer function to that of the interstellar medium.
The signal processing can be done either on-line using finite
impulse response filter devices \cite{bdz+97} or off-line in 
software \cite{sta98,sst+00}. The on-line approach allows for large
bandwidths to be employed and real-time viewing of the data. 
Off-line reduction, while slow and computationally expensive,
allows for more flexible data reduction schemes as well
as periodicity searches to be carried out.

The maximum time resolution obtainable via coherent dedispersion is the
inverse of the receiver bandwidth. For bandwidths of 10 MHz,
this technique makes it possible to resolve features on time-scales as
short as 100 ns. This corresponds to probing regions in the neutron star
magnetosphere as small as 30 m!

\subsection{The timing model}
\label{sec:tmodel}

Ideally, in order to model the rotational behaviour of the neutron
star, we require TOAs measured by an inertial observer. An observatory
located on Earth experiences accelerations with respect to the neutron
star due to the Earth's rotation and orbital 
motion around the Sun and is therefore not in
an inertial frame. To a very good approximation, the centre-of-mass of
the solar system, the solar system barycentre, can be regarded as an
inertial frame. It is standard practice \cite{hun71} to transform the
observed TOAs to this frame using a planetary ephemeris such as the
JPL DE200 \cite{sta82}.  The transformation is summarised as the
difference between barycentric ($\cal T$) and observed ($t$) TOAs:
\begin{equation}
\label{equ:bary}
{\cal T} - t =
\frac{\underline{r} . \hat{\underline{s}}}{c} +
\frac{(\underline{r} . \hat{\underline{s}})^2-|\underline{r}|^2}{2cd} +
\Delta t_{\rm rel} -
\Delta t_{\rm DM}.
\end{equation}
Here $\underline{r}$ is the position of the Earth with respect to the
barycentre, $\hat{\underline{s}}$ is a unit vector in the direction
towards the pulsar at a distance $d$, and $c$ is the speed of
light. The first term on the right hand side of this expression is the
light travel time from the Earth to the solar system barycentre. For
all but the nearest pulsars, the incoming pulses can be approximated
by plane wavefronts. The second term, which represents the delay due
to spherical wavefronts and which yields the trigonometric parallax
and hence $d$, is presently only measurable for four nearby
millisecond pulsars \cite{ktr94,cfw94,sbm+97}. The term $\Delta t_{\rm
rel}$ represents the Einstein and Shapiro corrections due to general
relativistic effects within the solar system \cite{bh86}.  Since
measurements are often carried out at different observing frequencies
with different dispersive delays, the TOAs are generally referred to
the equivalent time that would be observed at infinite frequency.
This transformation corresponds to the term $\Delta t_{\rm DM}$ and
may be calculated from Equation \ref{equ:defdt}.

Following the accumulation of about ten to twenty barycentric TOAs
from observations spaced over at least several months, a surprisingly
simple model can be applied to the TOAs and optimised so that it is
sufficient to account for the arrival time of any pulse emitted during
the time span of the observations and predict the arrival times of
subsequent pulses. The model is based on a Taylor expansion of the
angular rotational frequency $\Omega = 2 \pi/P$ about a model value
$\Omega_{\circ}$ at some reference epoch $t_{\circ}$.  The model pulse
phase $\phi$ as a function of barycentric time is thus given by:
\begin{equation}
\label{equ:phi}
\phi({\cal T}) = \phi_{\circ} + ({\cal T} - {\cal T}_{\circ})
\Omega_{\circ} + \frac{1}{2} ({\cal T} - {\cal T}_{\circ})^2 \dot{\Omega}_{\circ}
+ \cdots,
\end{equation}
where $\phi_{\circ}$ is the pulse phase at ${\cal T}_{\circ}$. Based
on this simple model, and using initial estimates of the position,
dispersion measure and pulse period, a ``timing residual'' is
calculated for each TOA as the difference between the observed and
predicted pulse phases.

A set of timing residuals for the nearby pulsar B1133+16
spanning almost 10 years is shown for
illustrative purposes in Fig.~\ref{fig:1133}.
Ideally, the residuals should have a zero mean and be free from any
systematic trends (Fig.~\ref{fig:1133}a). Inevitably, however, due to
our {\it a-priori} ignorance of the rotational parameters, the model
needs to be refined in a bootstrap fashion.  Early sets of residuals
will exhibit a number of trends indicating a systematic error in one
or more of the model parameters, or a parameter not initially
incorporated into the model.
\begin{figure}[hbt]
\setlength{\unitlength}{1in}
\begin{picture}(0,3.2)
\put(0.5,3.7){\includegraphics{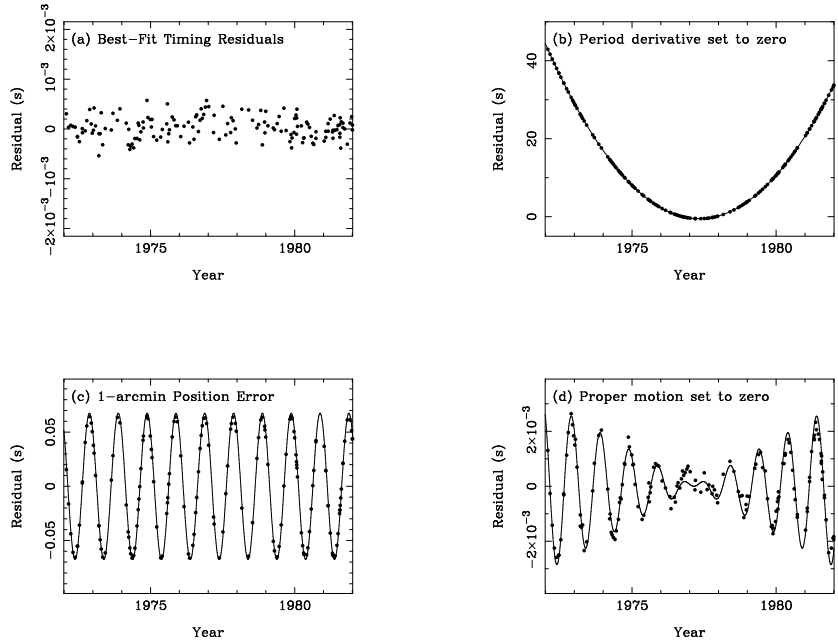}}
\end{picture}
\caption[]
{\sl
Timing model residuals versus date for PSR B1133+16. Case (a) shows
the residuals obtained from the best fitting model which includes
period, the period derivative, position and proper motion. Case (b) is the
result of setting the period derivative term to zero in this model.
Case (c) shows the effect of a 1 arcmin error in the assumed
declination.  Case (d) shows the residuals obtained assuming zero
proper motion. The lines in (b)--(d) show the expected behaviour
in the residuals for each effect (see text).
}
\label{fig:1133}
\end{figure}
From Equation \ref{equ:phi}, an error in the assumed $\Omega_{\circ}$
results in a linear slope with time.  A parabolic trend results from
an error in $\dot{\Omega}_{\circ}$ (Fig.~\ref{fig:1133}b).  Additional
effects will arise if the assumed position of the pulsar (the unit
vector $\hat{\underline{s}}$ in equation \ref{equ:bary}) used in the
barycentric time calculation is incorrect. A position error of just
one arcsecond results in an annual sinusoid (Fig.~\ref{fig:1133}c)
with a peak-to-peak amplitude of about 5 ms for a pulsar on the
ecliptic; this is easily measurable for typical TOA uncertainties of
order one milliperiod or better. A proper
motion produces an annual sinusoid of linearly increasing magnitude
(Fig.~\ref{fig:1133}d).

After a number of iterations, and with the benefit of a modicum of
experience, it is possible to identify and account for each of these
various effects to produce a ``timing solution'' which is phase
coherent over the whole data span. The resulting model parameters
provide spin and astrometric information about the neutron star to a
precision which improves as the length of the data span
increases.  Timing observations of the original
millisecond pulsar, B1937+21, spanning almost 9 yr (exactly
165,711,423,279 rotations!) measure a period of
$1.5578064688197945\pm0.0000000000000004$ ms \cite{ktr94,kas94}
defined at midnight UT on
December 5 1988!  Astrometric measurements based on these data are no
less impressive, with position errors of $\sim 20 \, \mu$arcsec being
presently possible.

\subsection{Timing stability}
\label{sec:tstab}

Ideally, after correctly applying a timing model,
we would expect a set of uncorrelated timing residuals
scattered in a Gaussian fashion about a zero mean with an rms
consistent with the measurement uncertainties. This is not always the
case; the residuals of many pulsars exhibit a quasi--periodic
wandering with time.
\begin{figure}[hbt]
\setlength{\unitlength}{1in}
\begin{picture}(0,3.5)
\put(1.5,-0.3){\includegraphics{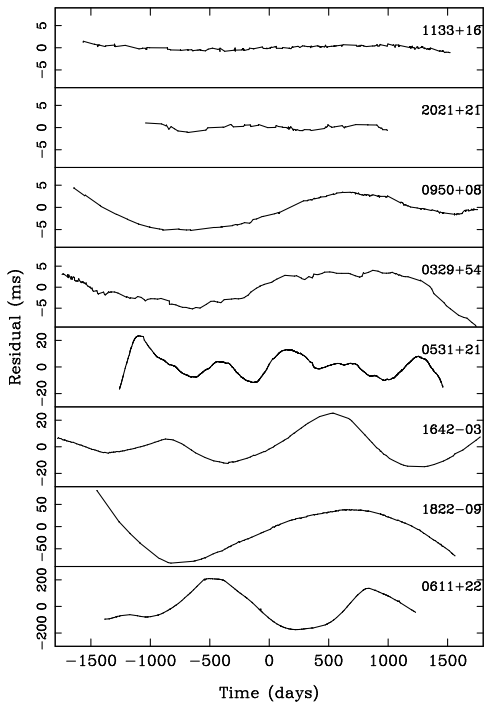}}
\end{picture}
\caption[]
{\sl
Examples of timing residuals for a number of normal pulsars. Note the
varying scale on the ordinate axis, the pulsars being ranked in
increasing order of timing ``activity''.
}
\label{fig:tnoise}
\end{figure}
A number of examples are shown in Fig.~\ref{fig:tnoise}. These are
taken from the Jodrell Bank timing program \cite{sl96}.  Such
``timing noise'' is most prominent in the youngest of the normal
pulsars \cite{mt74,ch80} and virtually absent in the much older
millisecond pulsars \cite{ktr94}.  While the physical processes of
this phenomena are not well understood, it seems likely that it
may be connected to superfluid processes and temperature changes
in the interior of the neutron star \cite{anp86} or processes in the
magnetosphere \cite{che87a,che87b}.

The relative dearth of timing noise for the older pulsars is a very
important finding. It implies that, presently, the measurement
precision depends primarily on the particular hardware constraints of
the observing system. Consequently, a large effort in hardware
development is presently being made to improve the precision of these
observations using, in particular, coherent dedispersion outlined in
\S \ref{sec:timobs}. Much of the pioneering work in this area has been
made by Joseph Taylor and collaborators at Princeton University
\cite{pripsr}. From high quality observations made using the Arecibo
radio telescope spanning almost a decade \cite{rt91a,rt91b,ktr94}, the
group has demonstrated that the timing stability of millisecond
pulsars over such time-scales is comparable to terrestrial atomic
clocks.

\begin{figure}[hbt]
\setlength{\unitlength}{1in}
\begin{picture}(2,2.8)
\put(2.0,0.0){\includegraphics{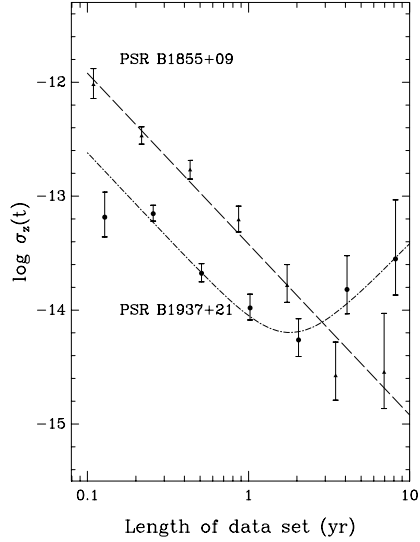}}
\end{picture}
\caption[] 
{\sl 
Fractional timing instabilities for PSRs B1855+09 and B1937+21 
as a function of time.  (After Kaspi, Taylor \& Ryba 1994 \cite{ktr94}).
}
\label{fig:mallen}
\end{figure}

This phenomenal stability is demonstrated in Fig.~\ref{fig:mallen}.
This figure shows $\sigma_z$, a parameter closely resembling the Allan
variance used by the clock community to estimate the stability of
atomic clocks \cite{tay91,allan}. Atomic clocks are known to have $\sigma_z
\sim 5 \times 10^{-15}$ on time-scales of order 5 years. The timing
stability of PSR B1937+21 seems to be limited by a power law component
which produces a minimum in its $\sigma_z$ after $\sim 2$ yr. This is
most likely a result of a small amount of intrinsic timing noise
\cite{ktr94}.  No such noise component is observed for
PSR B1855+09. This demonstrates that the timing stability
for PSR B1855+09 becomes competitive with the atomic clocks after about
3 yr.  The absence of timing noise for B1855+09 is probably related to
its characteristic age $\sim 5$ Gyr which is about a factor of 20
larger than B1937+21. Timing observations of millisecond
pulsars are discussed further
in the context of the pulsar timing array in \S \ref{sec:array}.

\subsection{Binary pulsars and Kepler's laws}
\label{sec:tbin}

For binary pulsars, the simple timing model introduced in
\S \ref{sec:tmodel} needs to be extended to incorporate the
additional radial acceleration of the pulsar as it orbits the common
centre-of-mass of the binary system.  Treating the binary orbit using
Kepler's laws to refer the TOAs to the binary barycentre requires
five additional model parameters: the orbital period ($P_b$),
projected semi-major orbital axis ($a_p \sin i$, see below), 
orbital eccentricity ($e$), longitude of periastron ($\omega$)
and the epoch of periastron passage ($T_0$). This description, using
five ``Keplerian parameters'', is identical to that used for
spectroscopic binary stars.

For spectroscopic binaries the orbital velocity curve shows the radial
component of the star's velocity as a function of time. The analogous
plot for pulsars is the apparent pulse period against time.  Two
examples are given in Fig.~\ref{fig:orbits}.
\begin{figure}[hbt]
\setlength{\unitlength}{1in}
\begin{picture}(0,1.9)
\put(-0.1,2.1){\includegraphics{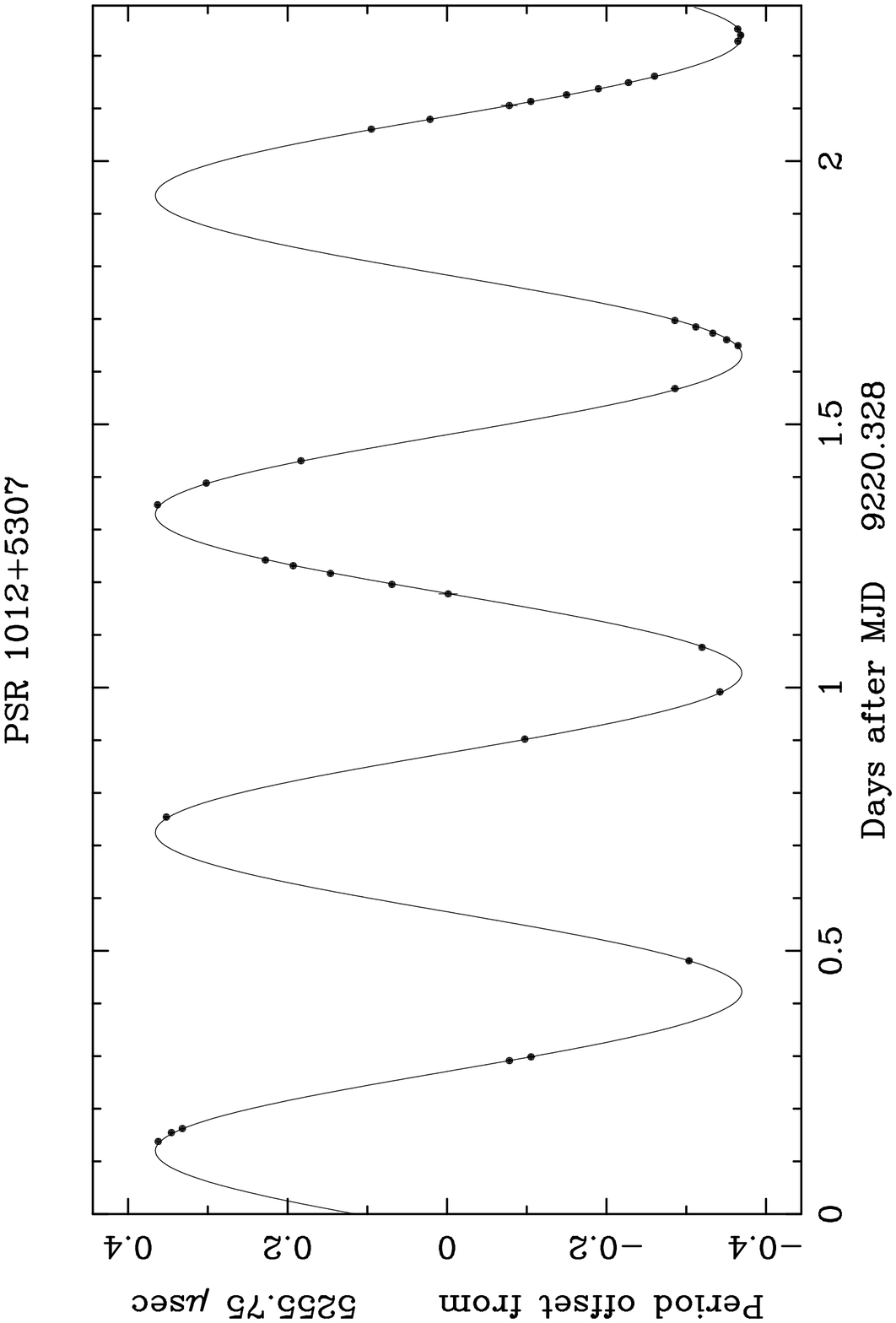}}
\put(+3.1,2.1){\includegraphics{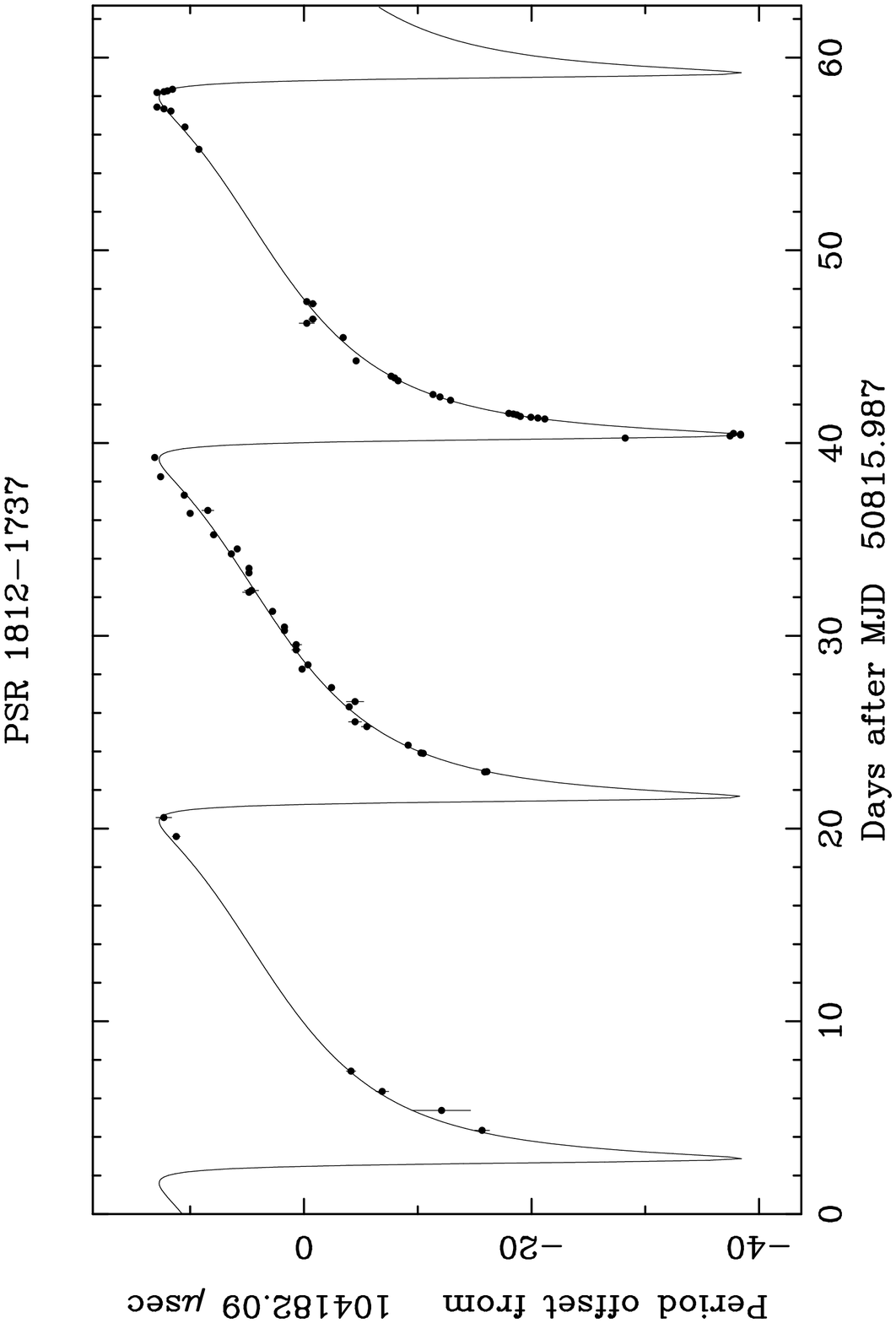}}
\end{picture}
\caption[] 
{\sl
Orbital velocity curves for two binary pulsars. Left: PSR
J1012$+$5307, a 5.25-ms pulsar in a 14.5-hour circular orbit around a
low-mass white dwarf companion \cite{nll+95,vk99,lcw+01}. Right: PSR
J1811$-$1736, a 104-ms pulsar in a highly eccentric 18.8-day orbit
around a massive companion (probably another neutron star) \cite{lcm+00}.
}
\label{fig:orbits}
\end{figure}
Constraints on the mass
of the orbiting companion can be placed by combining the projected
semi-major axis $a_{\rm p}\,\sin\,i$ and the orbital period 
to obtain the mass function:
\begin{equation}
\label{equ:massfn}
f(m_{\rm p},m_{\rm c}) = \frac{4\pi^2}{G} \frac{(a_{\rm
p}\,\sin\,i)^3}{P_b^2} = \frac{(m_{\rm c}\,\sin\,i)^3}{(m_{\rm
p}+m_{\rm c})^2},
\end{equation}
where $G$ is the universal gravitational constant.  Assuming a pulsar
mass $m_{\rm p}$ of 1.35 M$_{\odot}$ (see below), the mass of the
orbiting companion $m_{\rm c}$ can be estimated as a function of the
(initially unknown) angle $i$ between the orbital plane and the plane
of the sky. The minimum companion mass $m_{\rm min}$ occurs when the
orbit is assumed edge-on ($i=90^{\circ}$). For a random distribution of
orbital inclination angles, the probability of observing a binary
system at an angle {\it less} than some value $i_o$ is $p(<i_o) = 1 -
\cos(i_o)$. This implies that the chances of observing a binary system
inclined at an angle $\lapp$ 26$^{\circ}$ is only 10\%; evaluating the
companion mass for this inclination angle $m_{90}$ constrains the mass
range between $m_{\rm min}$ and $m_{90}$ at the 90\% confidence level.

\subsection{Post-Keplerian parameters}
\label{sec:postkep}

Although many of the presently known binary pulsar systems can be
adequately described by a Keplerian orbit, there are several
systems, including the original binary pulsar B1913+16, which exhibit
relativistic effects that require an additional set of
up to five ``post-Keplerian'' parameters. Within the framework of
general relativity, these can be written \cite{bt76} as:
\begin{eqnarray}
\dot\omega &=& 3 \left(\frac{P_b}{2\pi}\right)^{-5/3}
  (T_\odot M)^{2/3}\,(1-e^2)^{-1}\,, \label{equ:omdot} \\
\gamma &=& e \left(\frac{P_b}{2\pi}\right)^{1/3}
  T_\odot^{2/3}\,M^{-4/3}\,m_c\,(m_p+2m_c) \,, \\
\dot P_b &=& -\,\frac{192\pi}{5}
  \left(\frac{P_b}{2\pi}\right)^{-5/3}
  \left(1 + \frac{73}{24} e^2 + \frac{37}{96} e^4 \right)
  (1-e^2)^{-7/2}\,T_\odot^{5/3}\, m_p\, m_c\, M^{-1/3}\,,
  \label{equ:pbdot} \\
r &=& T_\odot\, m_c\,, \label{equ:r}\\
s &=& x \left(\frac{P_b}{2\pi}\right)^{-2/3}
  T_\odot^{-1/3}\,M^{2/3}\,m_c^{-1}\,. \label{equ:s}
\end{eqnarray}
In addition to the symbols defined above for Equation
\ref{equ:massfn}, $M\equiv m_p+m_c$, $x\equiv a_p \sin i/c$,
$s\equiv\sin i$ and $T_\odot\equiv GM_\odot/c^3 \simeq 4.925\,\mu$s.
All masses are in solar units. 

Measurements of post-Keplerian parameters for PSR B1913+16 have
been carried out by Taylor and a number of collaborators over the
years with steadily improving precision. The first of these parameters
to be measured was the advance of the longitude of periastron
($\dot{\omega}$).  This measurement is analogous to the perihelion
advance of Mercury \cite{mw75}. For PSR B1913+16 this amounts to about 
4.2 degrees per year \cite{thf+76}, some 4.6 orders of magnitude larger 
than for Mercury. A measurement of $\dot{\omega}$ alone yields the total 
mass for this system, $M=2.83$ M$_{\odot}$ assuming this advance
is due to general relativity.

Measurement of a second post-Keplerian parameter for B1913+16, $\gamma$, 
(gravitational redshift and transverse Doppler shifts in the orbit) permits an
unambiguous determination of $m_p$, $m_c$ and $i$ when combined with
$\dot{\omega}$ and the five Keplerian parameters. The original
measurements \cite{tfm79} have since
been substantially refined \cite{tw82,tw89} and 
the mass of the pulsar and its unseen companion have been determined to be 
$1.442\pm0.003$ M$_{\odot}$ and $1.386\pm0.003$ M$_{\odot}$ respectively.
Such phenomenal precision is a testament to the timing stability
of radio pulsars as clocks, and the diligence of Taylor and collaborators
in carrying out these long-term measurements.
Similar mass measurements now exist for the two other double neutron
star binary systems discussed in \S \ref{sec:nsns}: B1534+12
\cite{sac+98} and B2127+11C \cite{dk96}. For a number of other
systems, $\dot{\omega}$ measurements allow interesting constraints to
be placed on the component masses \cite{tamt93,nst96,tc99}.

An important general relativistic prediction for eccentric double
neutron star systems is the orbital decay due to the emission of
gravitational radiation ($\dot{P_b}$ in Equation
\ref{equ:pbdot}). Taylor et al.~\cite{tfm79,tw82,tw89} were able
to measure this for B1913+16 and found it to be in excellent agreement
with the predicted value.
\begin{figure}[hbt]
\setlength{\unitlength}{1in}
\begin{picture}(0,2.8)
\put(1.5,-0.4){\includegraphics{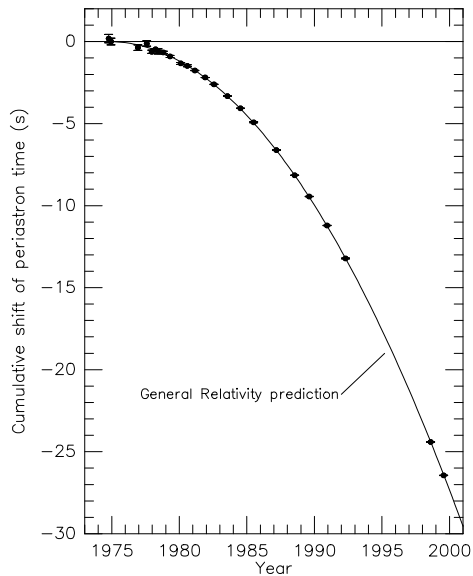}}
\end{picture}
\caption[]
{\sl
Orbital decay in the binary pulsar B1913+16 system demonstrated as an
increasing orbital phase shift for periastron passages with time. The general
relativistic prediction due entirely to the emission of gravitational
radiation is shown by the parabola.  
}
\label{fig:1913}
\end{figure}
The orbital decay, which corresponds to a shrinkage of about 3.2 mm
per orbit, is seen most dramatically as the gradually increasing shift
in orbital phase for periastron passages with respect to a
non-decaying orbit shown in Fig.~\ref{fig:1913}. This figure includes
recent Arecibo data taken in 1998 and 1999 following the upgrade of
the telescope in the mid 1990s. The observations of the orbital decay,
now spanning a 25-year baseline, are in agreement with general
relativity at the level of about 0.5\% and provide the first
(indirect) evidence for the existence of gravitational radiation.
Hulse and Taylor were awarded the Nobel prize in Physics in 1993
\cite{nobpr1993,hul94,tay94} in recognition of their
discovery of this remarkable laboratory for testing general relativity.

For those binary systems which are oriented nearly edge-on to the
line-of-sight, a significant delay is expected for orbital phases
around superior conjunction where the pulsar radiation is bent in the
gravitational potential well of the companion star. The so-called
``range'' and ``shape'' of the Shapiro delay effect are parameterized
by the last two post-Keplerian parameters $r$ and $s\equiv\sin i$ that
were introduced in Equations \ref{equ:r} and \ref{equ:s}. This effect,
analogous to the solar system Shapiro delay, has so far been
measured for two neutron star-white dwarf binary systems: B1855+09
and J1713+0747 \cite{rt91a,ktr94,cfw94} and for the double neutron star
binaries B1534+12 \cite{sac+98} and B1913+16 \cite{tw89}.

\subsection{Geodetic precession}
\label{sec:geodetic}

Shortly after the discovery of PSR B1913+16 it was realized that, if
the spin axis of the visible pulsar was misaligned with the angular
momentum axis of the binary system, the perturbing effect of the
companion on the space-time around the radio pulsar would cause it to
precess around the angular momentum axis \cite{dr74,eh75}. Within the
framework of general relativity, the rate of precession $\Omega_p$ was
shown \cite{bo75} to be
\begin{equation}
\Omega_p = 
\frac{(2\pi)^{5/3} \, T_{\odot}^{2/3} m_c \, (4 m_p + 3 m_c)}
     { P_{\rm b}^{5/3} \, (m_p+m_c)^{4/3}\, (1-e^2)},
\end{equation}
where we assume the same notation used for the discussion in \S
\ref{sec:tbin} and \S \ref{sec:postkep}. Inserting the parameters
of PSR B1913+16 yields $\Omega_p=1.21$ deg yr$^{-1}$. The period
of the precession is 297.5 yr. The 
observational consequence of geodetic precession is a secular
change in the pulse profile as the line-of-sight cut through
the emission beam changes (recall Fig.~\ref{fig:shapes}).

\begin{figure}[hbt]
\setlength{\unitlength}{1in}
\begin{picture}(0,2.8)
\put(0,3.6){\includegraphics{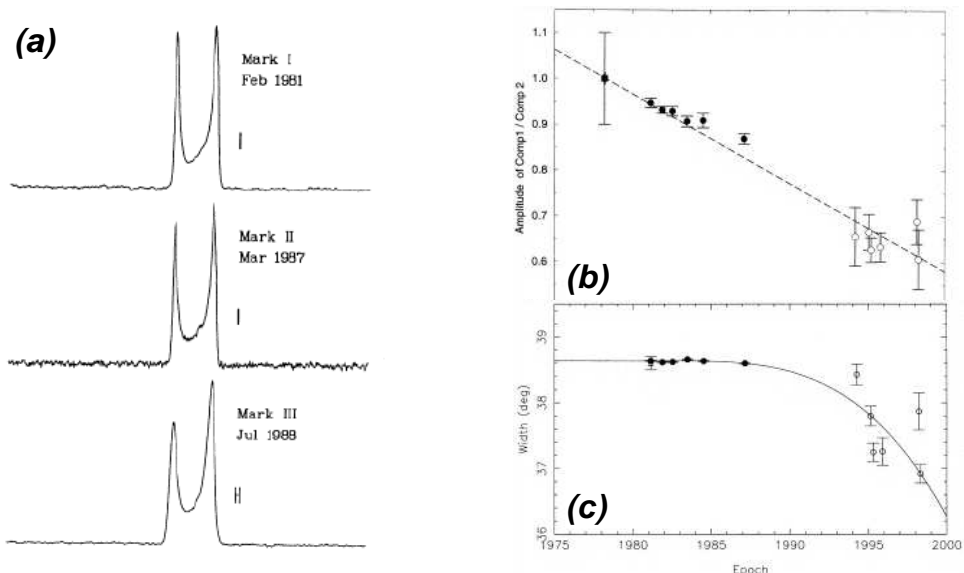}}
\end{picture}
\caption[]
{\sl
Geodetic precession in the binary pulsar B1913+16 system: (a) changes
in the observed pulse shapes between 1981--1988 seen as a decrease in
amplitude between the left and right components in the pulse profile
\cite{tw89,wrt89}; (b) relative heights of the two amplitudes plotted
between 1975--1998 \cite{kra98}; (c) component separation change \cite{kra98}.
}
\label{fig:1913geo}
\end{figure}

Early qualitative evidence for profile evolution due to
this effect \cite{tfm79} was substantiated with long-term Arecibo
measurements of component changes by Weisberg et al.~\cite{wrt89}.
Further changes were seen by Kramer with new Effelsberg data 
acquired in the 1990s \cite{kra98}. In addition to relative
amplitude variations, the expected changes in component separation 
for a hollow-cone beam model were also seen in the Effelsberg data. 
These observations are summarized in Fig.~\ref{fig:1913geo}.

In addition to the above results, there is now evidence for geodetic
precession in the other classic neutron star binary, PSR B1534+12
\cite{aptw96,stta00}. Although geodetic precession in binary pulsars
is another successful test of general relativity (albeit at a lower
precision than e.g.~orbital decay measurements), 
what is perhaps more interesting are the
various consequences it has. 
Geodetic precession only occurs when the spin and orbital
axes are misaligned \cite{bo75}. This is most likely to occur 
if the neutron star received an impulsive ``kick'' velocity
at birth (\S \ref{sec:pvel}). Wex et al.~\cite{wkk00} have
investigated the B1913+16 observations and find that the
kick magnitude was at least 250 km s$^{-1}$ and was directed almost
perpendicular to the spin axis of the neutron star progenitor.
This places stringent constraints on any kick mechanism.
Detailed monitoring of the pulse profile and polarization
properties now underway \cite{wt00,kra01} will allow the first
map of the emission beam of a neutron star to be made. This has
important implications for the various beaming models described
in \S \ref{sec:beaming}. There 
are already indications that the beam is circular \cite{kra01}.

The current results predict that B1913+16 will completely
precess out of the line of sight by around 2025 and re-appear
some 240 years later \cite{kra98}. Although we shall lose a most treasured
pulsar, we can take comfort from the fact that other pulsars
will precess into our field of view. Perhaps one example is
the newly-discovered relativistic binary J1141$-$6545 \cite{klm+00}
discussed in \S \ref{sec:plane} and \S \ref{sec:nswd}. This
relatively bright object was apparently missed by two previous 
searches during the early 1990s \cite{jlm+92,mld+96,lml+98}.

\subsection{Going further}
\label{sec:tfurther}

This chapter has outlined past and present progress in a number of
areas related to pulsar timing.  For further details on the technical
details and prospects of pulsar timing, the interested reader is
referred to a number of excellent review articles
\cite{wil81,bh86,tay91,bel98,bel98c}.  Two freely available
software packages which are routinely used for time-of-arrival
analyses by the pulsar community are available {\it viz:}~{\sc TEMPO}
\cite{tw89,pripsr} and {\sc TIMAPR}
\cite{dk95,timapr}. These packages are based on more detailed versions
of the timing model outlined in \S \ref{sec:tmodel}. An up-to-date list
summarising the various timing programmes is kept by Don Backer
\cite{bkypsr}. An audio file and slides from a lecture on pulsar
timing presented by Backer at the centennial meeting of the American
Physical society is also available on-line
\cite{backertalk}. Another relevant lecture from that meeting is
Will's presentation \cite{willtalk} on tests of Einstein's relativity
which includes an excellent overview of Taylor and Weisberg's
measurements of PSR B1913+16. Kramer \cite{kra01} has written a lucid
review article discussing measurements of geodetic precession in
binary pulsars and their implications.

The remarkable precision of these measurements, particularly for
millisecond pulsars, allows the detection of radial accelerations on
the pulsar induced by orbiting bodies smaller than the Earth. Alex
Wolszczan detected one such ``pulsar planetary system'' in 1990
following the discovery of a 6.2-ms pulsar B1257+12. In this case, the
pulsar is orbited by at least three Earth-mass bodies
\cite{wf92,psrplanets,wdk+00}.  Subsequent measurements of B1257+12 were even
able to measure resonance interactions between two of the planets
\cite{wol94}, confirming the nature of the system beyond all doubt.
Long-term timing measurements of the 11-ms pulsar B1620--26 in the
globular cluster M4 indicate that it may also have a planetary
companion \cite{tat93,bfs93,tacl99}. For detailed reviews of these systems,
and their implications for planetary formation scenarios, the
interested reader is referred to \cite{ptk93}.

\clearpage
\section{Pulsars as Gravitational Wave Detectors}
\label{sec:gwdet}

Many cosmological models predict that the Universe is presently filled
with a stochastic gravitational wave background (GWB) produced
during the big bang era \cite{pee93}.  The idea to use pulsars as
natural detectors of gravitational waves was first explored
independently by both Sazhin and Detweiler in the late 1970s
\cite{saz78,det79}.  The basic concept is to treat the solar system
barycentre and a distant pulsar as opposite ends of an imaginary arm
in space. The pulsar acts as the reference clock at one end of the arm
sending out regular signals which are monitored by an observer on the
Earth over some time-scale $T$. The effect of a passing gravitational
wave would be to cause a change in the observed rotational frequency
by an amount proportional to the amplitude of the wave.  For regular
monitoring observations of a pulsar with typical TOA uncertainties of
$\epsilon_{\rm TOA}$, this ``detector'' would be sensitive to waves
with dimensionless amplitudes $\gapp \epsilon_{\rm TOA}/T$ and
frequencies as low as $\sim 1/T$ \cite{bcr83,bnr84}. This method, which
already yields interesting upper limits on the GWB, is reviewed
in \S \ref{sec:uplim}. The idea of more sensitive detector
based on an array of pulsar clocks distributed over the sky is
discussed in \S \ref{sec:array}.

\subsection{Limits from individual pulsars}
\label{sec:uplim}

In the ideal case, the change in the observed frequency caused by the
GWB should be detectable in the set of timing residuals after the
application of an appropriate model for the rotational, astrometric
and, where necessary, binary parameters of the pulsar.  As discussed
in \S \ref{sec:pultim}, all other effects being negligible, the rms scatter
of these residuals $\sigma$ would be due to the measurement
uncertainties and intrinsic timing noise from the neutron
star. Detweiler \cite{det79} showed that a GWB with a flat energy
spectrum in the frequency band $f \pm f/2$ would result in an
additional contribution to the timing residuals $\sigma_g$.  The
corresponding wave energy density $\rho_g$ (for $fT \gg 1$) is
\begin{equation}
\rho_g = \frac{243 \, \pi^3 \, f^4 \, \sigma_g^2}{208 \, G}.
\end{equation}
An upper limit to
$\rho_g$ can be obtained from a set of timing residuals by assuming
the rms scatter is entirely due to this effect ($\sigma=\sigma_g$).
These limits are commonly expressed as a fraction of $\rho_c$ the
energy density required to close the Universe:
\begin{equation}
\rho_c = \frac{3 \, H_{\circ}^2}{8 \pi \,G}
\simeq 2 \times10^{-29} \, h^2 \,\,{\rm g \, cm}^{-3},
\end{equation}
where the Hubble constant $H_{\circ} = 100 \, h$ km s$^{-1}$ Mpc.

Romani \& Taylor \cite{rt83} applied this technique
to a set of TOAs for PSR B1237+12 obtained from regular observations
over a period of 11 years as part of the JPL pulsar timing programme
\cite{dr83}. This pulsar was chosen on the basis of its relatively low
level of timing activity by comparison with the youngest pulsars,
whose residuals are ultimately plagued by timing noise (\S
\ref{sec:tstab}). By ascribing the rms scatter in the residuals
($\sigma =240$ ms) to the GWB, Romani
\& Taylor placed a limit of $\rho_g/\rho_c \lapp 4 \times
10^{-3} h^{-2}$ for a centre frequency $f = 7 \times 10^{-9}$ Hz.

This limit, already well below the energy density required to close
the Universe, was further reduced following the long-term timing
measurements of millisecond pulsars at Arecibo by Taylor and
collaborators (\S \ref{sec:tstab}). In the intervening period, more
elaborate techniques had been devised \cite{bcr83,bnr84,srtr90} to
look for the likely signature of a GWB in the frequency spectrum of
the timing residuals and to address the possibility of ``fitting
out'' the signal in the TOAs. Following \cite{bcr83} it is convenient
to define $\Omega_g$, the energy density of the GWB per logarithmic
frequency interval relative to $\rho_c$. With this definition, the
power spectrum of the GWB, ${\cal P}(f)$, can be written
\cite{hr84,bnr84} as
\begin{equation}
{\cal P}(f) = \frac{G \, \rho_g}{3\pi^3 \, f^4} =
\frac{H_{\circ}^2 \, \Omega_g}{8\pi^4 \, f^5} =
1.34 \times 10^4 \, \Omega_g h^2 \, f^{-5}_{\rm yr^{-1}}
\,\, \mu{\rm s}^2 \, {\rm yr},
\end{equation}
where $f_{\rm yr^{-1}}$ is frequency in cycles per year. The timing
residuals for B1937+21 shown in Fig.~\ref{fig:ktr94res} are clearly
non-white and, as we saw in \S \ref{sec:tstab}, limit its timing
stability for periods $\gapp 2$ yr.  The residuals for PSR B1855+09
clearly show no systematic trends and are in fact consistent with the
measurement uncertainties alone. Based on these data, and using a
rigorous statistical analysis, Thorsett \& Dewey \cite{td96} place a
95\% confidence upper limit of $\Omega_g h^2 < 10^{-8}$ for $f = 4.4
\times 10^{-9}$ Hz. This limit is difficult to reconcile with most
cosmic string models for galaxy formation \cite{ca92,td96}.

\begin{figure}[hbt]
\setlength{\unitlength}{1in}
\begin{picture}(0,2)
\put(-0.2,2.16){\includegraphics{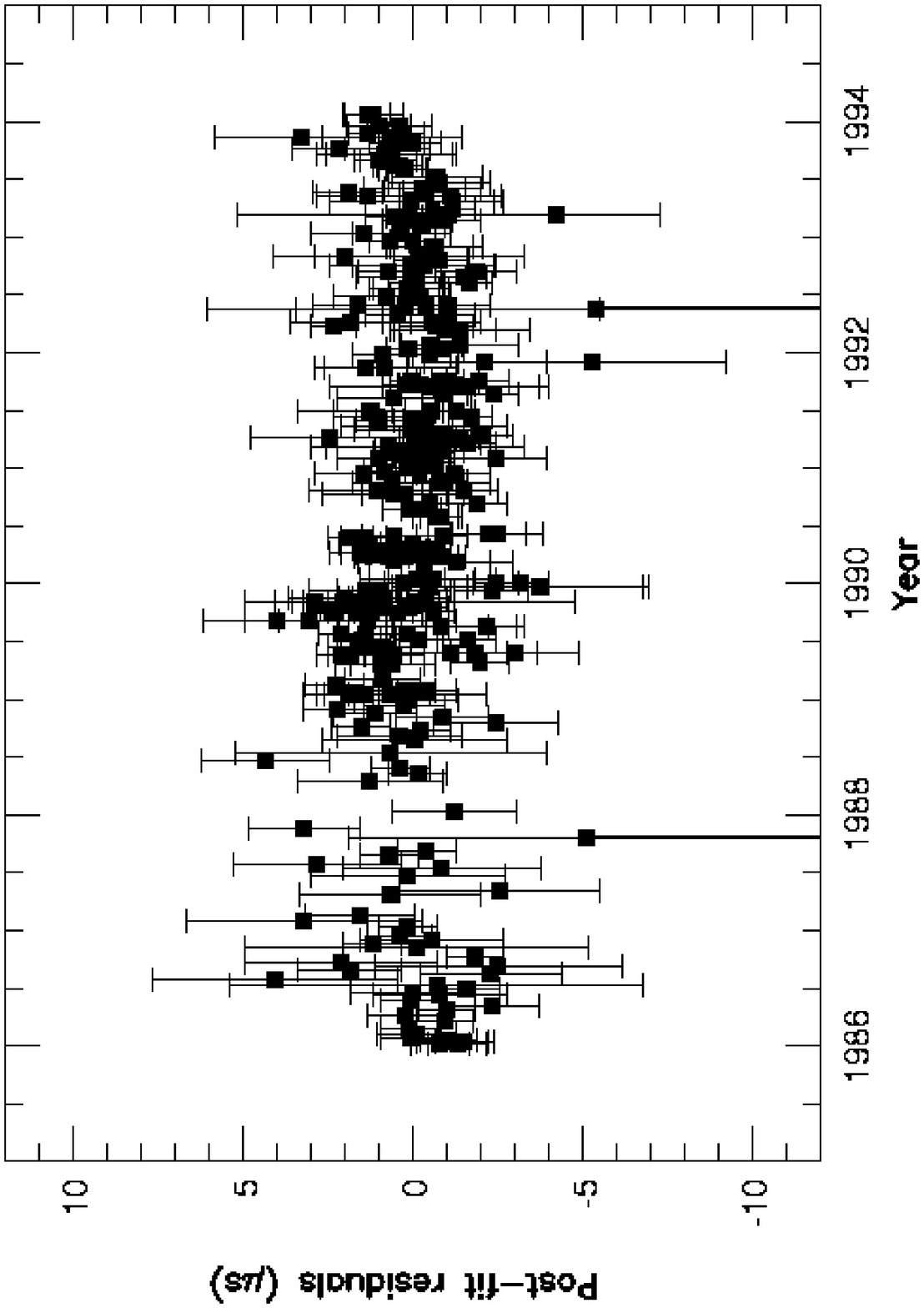}}
\put(+3.0,2.16){\includegraphics{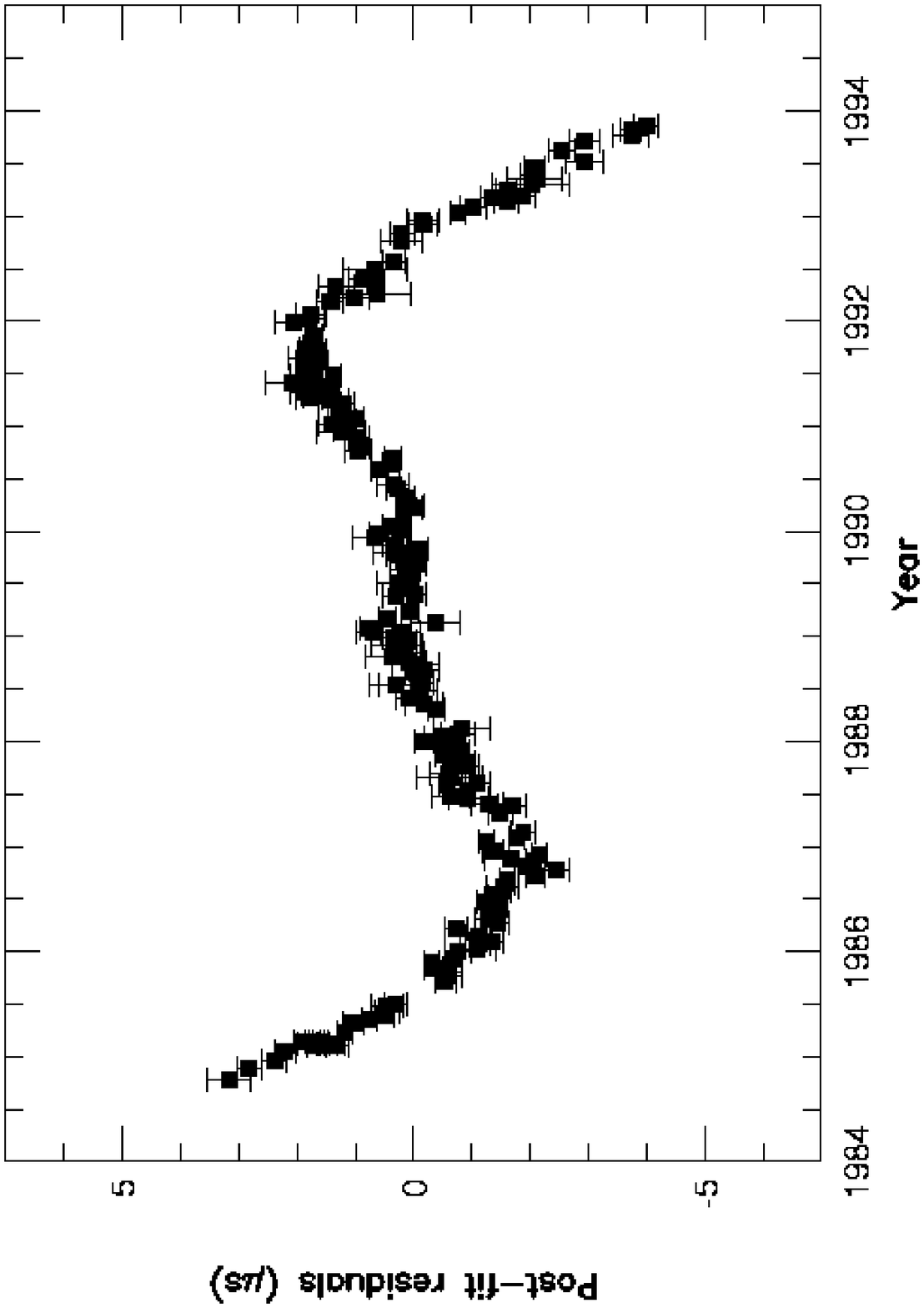}}
\end{picture}
\caption[]
{\sl
Timing residuals for PSRs B1855+09 (left panel) B1937+21 and (right panel)
obtained from almost a decade of timing at Arecibo. (Kaspi,
Taylor \& Ryba 1994 \cite{ktr94}).
}
\label{fig:ktr94res}
\end{figure}

For those pulsars in binary systems, an additional clock for measuring
the effects of gravitational waves is the orbital period.
In this case, the range of frequencies is not limited by the
time span of the observations, allowing the detection of waves with
periods as large as the light travel time to the binary system
\cite{bcr83}. The most stringent results presently available are based
on B1855+09 limit $\Omega_g h^2 < 2.7 \times 10^{-4}$ in the frequency
range $10^{-11} < f < 4.4 \times 10^{-9}$ Hz. Kopeikin \cite{kop97}
has recently presented this limit and discusses the methods in
detail.

\subsection{A pulsar timing array}
\label{sec:array}

The idea of using timing data for a number of pulsars distributed on
the sky to detect gravitational waves was first proposed by
Hellings \& Downs \cite{hd83}. Such a ``timing array'' of pulsars
would have the advantage over a single arm in that, through a
cross-correlation analysis of the residuals for pairs of pulsars distributed
over the sky, it should be possible to separate the timing noise of
each pulsar from the signature of the GWB, which would be
common to all pulsars in the array.
To quantify this, consider the fractional frequency shift observed for
the $i^{\rm th}$ pulsar in the array:
\begin{equation}
\frac{\delta \nu_i}{\nu_i} = \alpha_i {\cal A}(t) + {\cal N}_i(t).
\end{equation}
In this expression $\alpha_i$ is a geometric factor dependent on
the line-of-sight direction to the pulsar and the propagation
and polarisation vectors of the gravitational wave of dimensionless
amplitude ${\cal A}$. The timing noise intrinsic to the pulsar
is characterised by the function ${\cal N}_i$. The result of a
cross-correlation between pulsars $i$ and $j$ is then
\begin{equation}
\alpha_i \alpha_j <{\cal A}^2>
+ \alpha_i <{\cal A}{\cal N}_j>
+ \alpha_j <{\cal A}{\cal N}_i>
+ <{\cal N}_i{\cal N}_j>,
\end{equation}
where the bracketed terms indicate cross-correlations. Since the wave
function and the noise contributions from the two pulsars are
independent quantities, the cross correlation tends to $\alpha_i
\alpha_j <{\cal A}^2>$ as the number of residuals becomes
large. Summing the cross-correlation functions over a large number of
pulsar pairs provides additional information on this term as a
function of the angle on the sky \cite{hel90}. This allows
the separation of the effects of terrestrial clock and solar system
ephemeris errors from the GWB \cite{fb90}.

Applying the timing array concept to the {\it present} database of
long-term timing observations of millisecond pulsars does not improve
on the limits on the GWB discussed above. The sky distribution of
these pulsars, seen in the left panel of Fig.~\ref{fig:mspait}, shows
that their angular separation is rather low.  To achieve optimum
sensitivity it is desirable to have an array consisting of pulsar
clocks distributed isotropically over the whole sky. The flood of
recent discoveries of nearby binary and millisecond pulsars 
has resulted in essentially such a distribution,
shown in the right panel of Fig.~\ref{fig:mspait}. 

\begin{figure}[hbt]
\setlength{\unitlength}{1in}
\begin{picture}(2,1.7)
\put(-0.2,2.8){\includegraphics{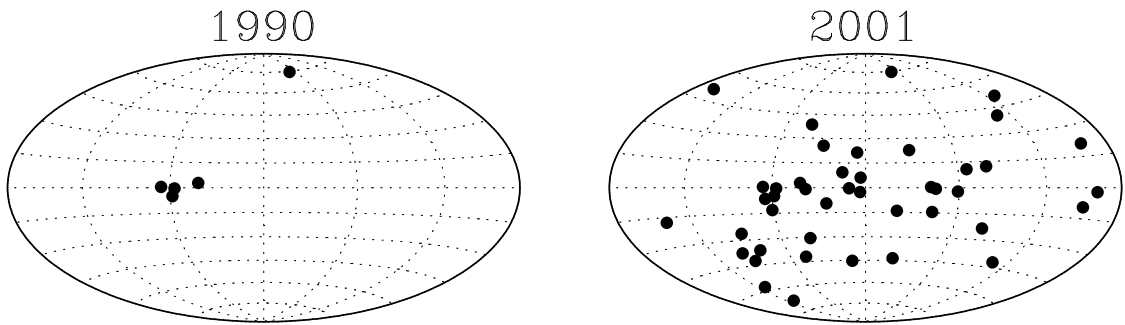}}
\end{picture}
\caption[] 
{\sl 
Hammer-Aitoff projections showing the known Galactic disk millisecond
pulsar population in 1990 and 2001. The impact of the new discoveries
is seen by comparing the sample circa 1990, where all the known
sources had been discovered at Arecibo, with the present sample where
the sources are much more uniformly distributed on the sky.
}
\label{fig:mspait}
\end{figure}

A number of long-term timing projects are now underway to monitor
these millisecond pulsars with a goal of detecting low-frequency
gravitational radiation. At Arecibo, regular timing of a dozen or more
millisecond pulsars has been carried out following the completion of the
upgrade to the telescope in 1997. A summary of these observations is
shown in Fig.~\ref{fig:ctiming}. The rms timing residuals for several
of the pulsars are now approaching the 100 ns level. This degree of
precision demands a high level of commitment to investigate possible
causes of systematic errors in the signal path through the
telescope. Combining datasets from several observatories is also
challenging. The Berkeley pulsar group lead by Don Backer
\cite{bkypsr} are among the most active observers in this area. Backer
and collaborators have now installed identical sets of datataking
equipment at a number of radio telescopes around the world in an
attempt to ensure a homogeneous set of residuals.  Continued timing of
these millisecond pulsars in the coming years should greatly improve the
sensitivity and will perhaps allow the detection of gravitational
waves, as opposed to upper limits, in the not-too-distant future.

\begin{figure}[hbt]
\setlength{\unitlength}{1in}
\begin{picture}(0,5)
\put(-0.1,5){\includegraphics{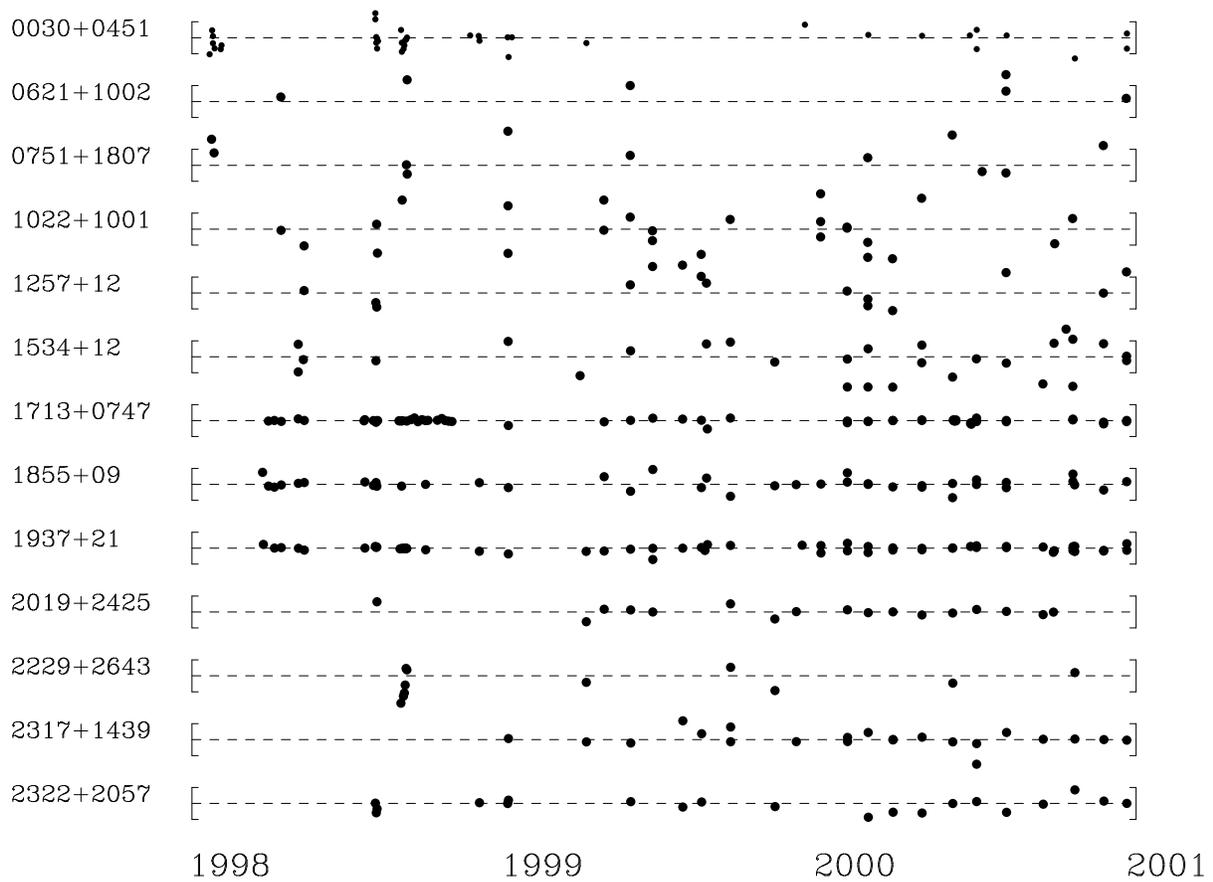}}
\end{picture}
\caption[]
{\sl
Arecibo timing residuals of millisecond pulsars monitored by groups at Berkeley
and Princeton on a regular basis as part of a long-term project to
detect low-frequency gravitational waves. The level of precision being
achieved is well below the vertical bars surrounding each set of residuals 
($\pm 5 \mu$s).
}
\label{fig:ctiming}
\end{figure}

\subsection{Going further}

Further discussions on the realities of using pulsars as gravity
wave detectors can be found in two excellent review articles by
Romani \cite{rom89} and Backer \cite{bac96}.

\clearpage
\section{Summary and Future Prospects}
\label{sec:future}

The main aim of this article was to review some of the many
astrophysical applications provided by the present sample of binary
and millisecond radio pulsars.  The topics covered here, along with the
bibliography and associated tables of observational parameters, should
be useful to those wishing to delve deeper into the vast body of literature
that exists. We now briefly recap on the main issues.

Through an understanding of the Galactic population of radio pulsars
summarised in \S \ref{sec:gal} it is possible to predict the detection
statistics of terrestrial gravitational wave detectors to nearby
rapidly spinning neutron stars (\S \ref{sec:nmsppop}), as well as
coalescing relativistic binaries at cosmic distances (\S
\ref{sec:relpop}).  Continued improvements in gravitational wave
detector sensitivities should result in a number of interesting
developments and contributions in this area. These developments and
contributions might include the detection of presently known radio
pulsars, as well as a population of coalescing binary systems which
have not yet been detected as radio pulsars.
The phenomenal timing stability of radio pulsars leads naturally to a
large number of applications, including their use as laboratories for
relativistic gravity (\S \ref{sec:postkep}) and as natural detectors
of gravitational radiation (\S \ref{sec:gwdet}). Long-term timing
experiments of the present sample of millisecond and binary pulsars
currently underway appear to have tremendous potential in these areas
and perhaps detect the gravitational wave background (if it exists) 
within the next decade.

These applications will benefit greatly from the continued
discovery of new systems by the present generation of radio pulsar
searches which continue to probe new areas of parameter
space. Based on the results presented in \S \ref{sec:nmsppop}, it is
clear that we are aware of only about 1\% of the total active pulsar
population in our Galaxy. It is therefore likely that we have
not seen all of the pulsar zoo. More sensitive surveys are being
planned both in the short term (a multibeam system on the Arecibo
telescope \cite{aomulti}) and in the longer term (the Square Kilometer 
Array \cite{ska}). These should provide a far more complete census 
of the Galactic pulsar
population. Possible discoveries in the future include:
\begin{itemize}
\item 
A dual-line binary pulsar, i.e.~a double neutron star system
in which both components are observable as radio pulsars. The additional
clock in such a binary system would be most valuable in further tests
of strong-field gravity.
\item 
A radio pulsar with a black-hole companion would undoubtably also
be a fantastic laboratory for studying gravity in the strong-field regime.
\item A sub-millisecond pulsar. The original millisecond pulsar,
B1937+21, rotating at 642 Hz is still the most rapidly rotating
neutron star known.  Do kHz neutron stars exist? Searches now have
sensitivity to such objects \cite{bd97} and a discovery of even one would
constrain the equation of state of matter at high densities.
\item 
A binary system in which the neutron star is in the process
of transforming from an X-ray-emitting neutron star to a millisecond 
radio pulsar. 
\end{itemize}
Spurred on by recent discoveries \cite{clf+00,lmbm00,rgh+01,dlm+01},
a number of high-sensitivity searches for pulsars in globular clusters are
being conducted. These have tremendous potential for discovering
new and exotic binary systems like a millisecond pulsar--black hole binary.

\clearpage
\subsection*{Acknowledgments}

Many thanks to Maura McLaughlin, Jiannis Seiradakis and Michael Kramer
who read and commented on earlier incantations of this revised review,
as well as a number of other colleagues who gave me useful feedback on
the original article. Jiannis Seiradakis urged me to include the
tables of parameters given in the appendix after I promised to put
them in the original article, but didn't. The tables and references
should be useful to both observers and theorists. Thanks also to
Fernando Camilo for allowing me to include details on a number of
pulsars in these tables prior to publication.

I am indebted to a number of colleagues who kindly gave permission to
use a selection of figures in this article. Michael Kramer provided
the cute animation of the rotating neutron star presented in
Fig.~\ref{fig:rotns}. Vicky Kalogera provided the graph
used in Fig.~\ref{fig:smallnumber}.
Joe Taylor and Joel Weisberg provided the
updated orbital decay curve of PSR B1913+16 shown in
Fig.~\ref{fig:1913}. Fig.~\ref{fig:tnoise}, based on unpublished
timing observations carried out at Jodrell Bank, was supplied by
Andrew Lyne. Vicky Kaspi provided the millisecond pulsar timing
residuals and comparison of their timing stabilities shown in
Figs.~\ref{fig:ktr94res} and \ref{fig:mallen} respectively.  Andrea
Lommen provided the pulsar timing residuals from the Arecibo timing
program used to produce Fig.~\ref{fig:ctiming}.  Frequent use was made
of NASA's magnificent Astrophysics Data System
\cite{nasaads} and LANL's preprint archives \cite{lanl}
during the literature searches.
Finally, I'd like to thank the Living
Reviews editor, Theresa Velden, for being extraordinarily patient, and
for putting up with many a feeble excuse from me during the writing
of this long-overdue update.

\appendix
\section{Tables of Binary and Millisecond Pulsars}

\begin{table}[hbt]
\footnotesize
\begin{center}
\begin{tabular}{lrrrrrr}
\hline
\hline
Name & $P$ & $\log(\tau_c)$ & $\log(B)$ & $d$ & $v_t$ & Ref. \\
     &   ms         &                   &                & kpc& km/s&  \\
\hline
J0030$+$0451 & 4.865  & 9.9 & 8.3  & 0.23 &$<$65& \cite{lzb+01} \\
J0711$-$6830 & 5.491  & 10.4& 8.2  & 1.04 & 139 & \cite{bjb+97,tsb+99} \\
J1024$-$0719 & 5.162  & 9.7 & 8.5  & 0.35 & 45  & \cite{bjb+97,tsb+99} \\
\\
J1730$-$2304 & 8.123  & 9.9 & 8.6  & 0.51 & 53  & \cite{lnl+95,tsb+99} \\
J1744$-$1134 & 4.075  & 9.9 & 8.3  & 0.17 & 20  & \cite{bjb+97,tsb+99} \\
B1937$+$21   & 1.558  & 8.4 & 8.6  & 9.65 & 22  & \cite{bkh+82,ktr94}  \\
\\
J2124$-$3358 & 4.931  & 9.9 & 8.4  & 0.25 & 67  & \cite{bjb+97,tsb+99}\\
J2235$+$1506 & 59.767 & 10.0 & 9.4 & 1.15 & 98  & \cite{cnt93} \\
J2322$+$2057 & 4.808  & 10.2 & 8.2 & 0.78 & 89  & \cite{nt95,cnt96} \\
\hline
\end{tabular}
\caption{\sl 
Parameters for the 9 isolated millisecond pulsars currently known in the
Galactic disk.
Listed are the spin period $P$, the base-10 logarithms of the characteric
age $\tau_c$ and surface magnetic field strength $B$ (\S \ref{sec:spinpars}),
the distance $d$ derived from the Taylor \& Cordes electron density model
\cite{tc93} or independently (when available) and the transverse speed
$v_t$ inferred from $d$ and a proper motion measurement (when available).
Key publications for each pulsar are referenced to the bibliography.
}
\label{tab:imsps}
\end{center}
\end{table}

\begin{table}[hbt]
\footnotesize
\begin{tabular}{lrrrrrrrrrr}
\hline
\hline
Name & $P$ & $\log(\tau_c)$ & $\log(B)$ & $d$ & $v_t$ & $P_b$ & $x$ & $e$ & $m_2$ & Ref. \\
     &   ms         &                   &                & kpc& km/s&
days & s & & M$_{\odot}$ & \\
\hline
J1141$-$6545 & 393.898 & 6.2 & 12.1 & 3.20 & ? & 0.20 & 1.86   & 0.17   & 1.0 & \cite{klm+00}\\
B1259$-$63 & 47.762 & 5.5 & 11.5 & 4.60 & ? & 1236.72 & 1296.58   & 0.87   & 10.0 & \cite{jlm+92,jml+92}\\
J1518$+$4904 & 40.935 & 10.3 & 9.0 & 0.70 & 27 & 8.63 & 20.04   & 0.25   & 1.3&\cite{nst96,nst99}\\
\\
B1534$+$12 & 37.904 & 8.4 & 10.0 & 0.68 & 80 & 0.42 & 3.73   & 0.27   & 1.3 &  \cite{wol91a,sac+98}\\
J1740$-$3052 & 570.309 & 5.5 & 12.6 & 10.8& ? & 231.03& 756.91& 0.58 & 16 & \cite{mlc+00}\\
J1811$-$1736 & 104.182 & 8.9 & 10.1& 5.94 & ? & 18.77 & 34.78 & 0.83 & 0.7 & \cite{lcm+00}\\
\\
B1820$-$11 & 279.828 & 6.5 & 11.8 & 6.26 & ? & 357.76 & 200.67   & 0.79   & 0.7& \cite{lm89,pv91}\\
B1913$+$16 & 59.030 & 8.0 & 10.4 & 7.13 & 100 & 0.32 & 2.34   & 0.62   & 1.4 &\cite{tw82,tw89}\\
B2303$+$46 & 1066.371 & 7.5 & 11.9 & 4.35 & ? & 12.34 & 32.69   & 0.66   & 1.2 & \cite{lb90,vk99}\\
\hline
\end{tabular}
\caption{\sl
Parameters for the 9 high-eccentricity ($e>0.15$)
binary pulsars currently known in the Galactic disk.
Listed are the spin period $P$, the base-10 logarithms of the characteric
age $\tau_c$ and surface magnetic field strength $B$ (\S \ref{sec:spinpars}),
the distance $d$ derived from the Taylor \& Cordes electron density model
\cite{tc93} or independently (when available), the transverse speed
$v_t$ inferred from $d$ and a proper motion measurement (when available),
the binary period $P_b$, the projected semi-major axis of the orbit $x$
in units of light seconds, the orbital eccentricity $e$ and the 
companion mass $m_2$ evaluated from the mass function assuming
a pulsar mass of 1.4 M$_{\odot}$ and an inclination angle of 60 degrees
(\S \ref{sec:tbin}) or (when known) from independent measurements. 
Key publications for each pulsar are referenced to the bibliography.
}
\label{tab:ebpsrs}
\end{table}

\begin{table}[hbt]
\footnotesize
\begin{tabular}{lrrrrrrrrrr}
\hline
\hline
Name & $P$ & $\log(\tau_c)$ & $\log(B)$ & $d$ & $v_t$ & $P_b$ & $x$ & $e$ & $m_2$ & Ref. \\
     &   ms         &                   &                & kpc& km/s&
days & s & & M$_{\odot}$ & \\
\hline
J0034$-$0534 & 1.877 & 9.9 & 7.9 & 0.98 & 71 & 1.59 & 1.44   & $<0.00002$  & 0.1 & \cite{bhl+94}\\
J0218$+$4232 & 2.323 & 8.7 & 8.6 & 5.85 & ? & 2.03 & 1.98   & $<0.00002$   & 0.2 & \cite{nbf+95}\\
J0437$-$4715 & 5.757 & 9.7 & 8.5 & 0.18 & 121 & 5.74 & 3.37   & 0.000019   & 0.1 & \cite{jlh+93}\\
J0613$-$0200 & 3.062 & 9.7 & 8.2 & 2.19 & 77 & 1.20 & 1.09   & 0.000007   & 0.1 &\cite{lnl+95,tsb+99}\\
J0621$+$1002 &28.854 & 10.1 & 9.0 & 1.88 & ? & 8.31  & 12.03  & 0.0025   &  0.5 & \cite{cnst96}\\
B0655$+$64 & 195.671 & 9.7 & 10.1 & 0.48 & 32 & 1.03 & 4.13   & 0.000008   & 0.7 & \cite{jl88,vk95}\\
J0751$+$1807 & 3.479 & 9.8 & 8.2 & 2.02 & ? & 0.26 & 0.40   & $>0.0$   & 0.1 & \cite{lzc95}\\
B0820$+$02 & 864.873 & 8.1 & 11.5 & 1.43 & 35 & 1232.47 & 162.15   & 0.011868   & 0.2& \cite{mncl80,vk95}\\
J1012$+$5307 & 5.256 & 9.8 & 8.4 & 0.52 & 102 & 0.60 & 0.58   & $<0.0000008$ & 0.1 & \cite{nll+95,lcw+01}\\
J1022$+$1001 & 16.453 & 9.8 & 8.9 & 0.60 & $>50$ & 7.81 & 16.77   & 0.000098   & 0.7 & \cite{cnst96,kxc+99}\\
\\
J1045$-$4509 & 7.474 & 10.0 & 8.5 & 3.25 & 52 & 4.08 & 3.02   & 0.000024   & 0.2&\cite{lnl+95,tsb+99}\\
J1157$-$5114 & 43.589 & 9.7 & 9.4 & 1.88 & ? & 3.51 & 14.29   & 0.00040   & 1.2 & \cite{eb01}\\
J1232$-$6501 & 88.282 & 9.2 & 9.9 & 10.00 & ? & 1.86 & 1.61  &  0.00011&  0.1 & \cite{clm+01}\\
B1257$+$12 & 6.219 & 9.5 & 8.6 & 0.62 & 281 & \multicolumn{4}{c}{planetary system} & \cite{wf92,wol94}\\
J1435$-$6100 & 9.348 & 9.8 & 8.7 & 3.25 & ? & 1.35 & 6.18 &0.00001 &  0.9& \cite{clm+01}\\
J1454$-$5846 & 45.249 & 9.0 & 9.8 & 3.32 & ? & 12.42  & 26.52 & 0.0019&  0.9 & \cite{clm+01}\\
J1455$-$3330 & 7.987 & 9.9 & 8.5 & 0.74 & 100 & 76.17 & 32.36   & 0.000170   & 0.3& \cite{lnl+95,tsb+99}\\
J1603$-$7202 & 14.842 & 10.2 & 8.7 & 1.64 & 27 & 6.31 & 6.88   & $>0.0$ & 0.3& \cite{llb+96,tsb+99}\\
J1640$+$2224 & 3.163 & 10.2 & 8.0 & 1.18 & 76 & 175.46 & 55.33   & 0.0008   & 0.3 & \cite{wdk+00} \\
J1643$-$1224 & 4.622 & 9.7 & 8.4 & 4.86 & 159 & 147.02 & 25.07   & 0.000506   & 0.1& \cite{lnl+95,tsb+99}\\
\\
J1713$+$0747 & 4.570 & 10.0 & 8.3 & 0.89 & 27 & 67.83 & 32.34   & 0.000075   & 0.3& \cite{fwc93,cfw94}\\
J1757$-$5322 & 8.870 & 9.7 & 8.7 & 1.36 & ? & 0.45 & 2.09   & 0.000004   & 0.6 & \cite{eb01}\\
B1800$-$27 & 334.415 & 8.5 & 10.9 & 3.62 & ? & 406.78 & 58.94   & 0.000507   & 0.1 & \cite{jlm+92}\\
J1804$-$2717 & 9.343 & 9.5 & 8.8 & 1.17 & ? & 11.13 & 7.28   & 0.000035   & 0.2 & \cite{llb+96}\\
J1810$-$2005 & 32.822 & 9.6 & 9.3 & 4.04 & ? & 15.01& 11.98& 0.000025&0.3 & \cite{clm+01}\\
B1831$-$00 & 520.954 & 8.8 & 10.9 & 2.63 & ? & 1.81 & 0.72   & $>0.0$   & 0.1 &\cite{dmr+86}\\
B1855$+$09 & 5.362 & 9.7 & 8.5 & 1.00 & 29.2 & 12.33 & 9.23   & 0.000022   & 0.2&\cite{srs+86,ktr94}\\
J1904$+$0412 & 71.095 & 10.1 & 9.4 & 4.01 & ? & 14.93 & 9.63& 0.00022&  0.2 & \cite{clm+01}\\
J1911$-$1114 & 3.626 & 0.0 & 0.0 & 1.59 & 183 & 2.72 & 1.76   & $>0.0$   & 0.1& \cite{llb+96,tsb+99}\\
B1953$+$29 & 6.133 & 9.5 & 8.6 & 5.39 & 98 & 117.35 & 31.41   & 0.00033   & 0.2 & \cite{bbf83,wdk+00}\\
\\
B1957$+$20 & 1.607 & 9.4 & 8.1 & 1.53 & 190 & 0.38 & 0.09   & $>0.0$   & 0.02&\cite{fst88,aft94}\\
J2019$+$2425 & 3.935 & 10.4 & 8.0 & 0.91 & 83 & 76.51 & 38.77   & 0.000111   & 0.3 & \cite{nt95,nss01}\\
J2033$+$1734 & 5.949 & 9.9 & 8.4 & 1.38 & ? & 56.31 & 20.16   & 0.00013   & 0.2&\cite{rtj+96}\\
J2051$-$0827 & 4.509 & 9.7 & 8.4 & 1.28 & 14 & 0.10 & 0.05   & $>0.0$   & 0.03&\cite{sbl+96,tsb+99}\\
J2129$-$5721 & 3.726 & 9.5 & 8.4 & 2.55 & 56 & 6.63 & 3.50   & $>0.0$   & 0.1&\cite{llb+96,tsb+99}\\
J2145$-$0750 & 16.052 & 10.3 & 8.6 & 0.50 & 38 & 6.84 & 10.16   & 0.000019   & 0.4 & \cite{bhl+94,tsb+99}\\
J2229$+$2643 & 2.978 & 10.4 & 7.9 & 1.43 & 113 & 93.02 & 18.91   & 0.00026   & 0.1 & \cite{wdk+00}\\
J2317$+$1439 & 3.445 & 10.6 & 7.9 & 1.89 & 68 & 2.46 & 2.31   & $>0.0$   & 0.2& \cite{cnt93,cam95a}\\
\hline
\end{tabular}
\caption{\sl
Parameters for 38 low-eccentricity binary pulsars currently known
in the Galactic disk.
Listed are the spin period $P$, the base-10 logarithms of the characteric
age $\tau_c$ and surface magnetic field strength $B$ (\S \ref{sec:spinpars}),
the distance $d$ derived from the Taylor \& Cordes electron density model
\cite{tc93} or independently (when available), the transverse speed
$v_t$ inferred from $d$ and a proper motion measurement (when available),
the binary period $P_b$, the projected semi-major axis of the orbit $x$
in units of light seconds, the orbital eccentricity $e$ (when measured)
and the companion mass $m_2$ evaluated from the mass function assuming
a pulsar mass of 1.4 M$_{\odot}$ and an inclination angle of 60 degrees
(\S \ref{sec:tbin}) or (when known) from independent measurements. 
Key publications for each pulsar are referenced to the bibliography.
}
\label{tab:bmsps}
\end{table}

\begin{table}[hbt]
\footnotesize
\begin{tabular}{lrlrrrrrr}
\hline
\hline
PSR & $P$&Cluster& $d$ & $P_b$& $x$ & $e$ & $m_2$ & Ref.\\
    &  (ms)  &         & (kpc)&  (days)       & (s)&&M$_{\odot}$& \\
\hline
J0023$-$7204C & 5.757 & 47~Tuc & 4.5 &                &    &   &    &
\cite{clf+00,fcl+01}\\
J0024$-$7204D & 5.358 & 47~Tuc & 4.5 &                &    &   &    & 
\cite{clf+00,fcl+01}\\
J0024$-$7205E & 3.536 & 47~Tuc & 4.5 & 2.26 & 1.98 & 0.0003  & 0.2& 
\cite{clf+00,fcl+01}\\
J0024$-$7204F & 2.624 & 47~Tuc & 4.5 &                &    &   &    &
\cite{clf+00,fcl+01}\\
J0024$-$7204G & 4.040 & 47~Tuc & 4.5 &                &    &   &    &
\cite{clf+00,fcl+01}\\
J0024$-$7204H & 3.210 & 47~Tuc & 4.5 & 2.36 & 2.15 & 0.07  & 0.2&
\cite{clf+00,fcl+01}\\
J0024$-$7204I & 3.485 & 47~Tuc & 4.5 & 0.23 & 0.04 & $<0.001$  & 0.01&
\cite{clf+00,fcl+01}\\
J0023$-$7203J & 2.101 & 47~Tuc & 4.5 & 0.12 & 0.04 & $<0.0002$ & 0.02&
\cite{clf+00,fcl+01}\\
J0024$-$7204L & 4.346 & 47~Tuc & 4.5 &                &    &   &    &
\cite{clf+00,fcl+01}\\
J0023$-$7205M & 3.677 & 47~Tuc & 4.5 &                &    &   &    &
\cite{clf+00,fcl+01}\\
J0024$-$7204N & 3.054 & 47~Tuc & 4.5 &                &    &   &    &
\cite{clf+00,fcl+01}\\
J0024$-$7204O & 2.643 & 47~Tuc & 4.5 & 0.14 & 0.05 & $<0.004$  & 0.02  &
\cite{clf+00,fcl+01}\\
J0024$-$72P & 3.643 & 47~Tuc & 4.5 & 0.14 & 0.04 &  $>0.0$      & 0.02 &
\cite{clf+00}\\
J0024$-$7204Q & 4.033 & 47~Tuc & 4.5 & 1.19 & 1.46 &  0.00007 & 0.2  &
\cite{clf+00,fcl+01}\\
J0024$-$72R & 3.480 & 47~Tuc & 4.5 & 0.066& 0.033&  $>0.0$      &0.03&
\cite{clf+00}\\
J0024$-$72S & 2.830 & 47~Tuc & 4.5 & 1.20  & 0.77&  $>0.0$      &0.09&
\cite{clf+00,fkl01}\\
J0024$-$7204T & 7.589 & 47~Tuc & 4.5 & 1.13 & 1.34 &  0.0004  &    & 
\cite{clf+00,fcl+01}\\
J0024$-$7203U & 4.343 & 47~Tuc & 4.5 & 0.43 & 0.53 &  0.0002 &    &
\cite{clf+00,fcl+01}\\
J0024$-$72V & 4.810 & 47~Tuc & 4.5 & 
\multicolumn{4}{c}{currently no orbital solution}&
\cite{clf+00}\\
J0024$-$72W & 2.352 & 47~Tuc & 4.5 & 0.11 & 0.24 &  $>0.0$      &0.15& 
\cite{clf+00}\\
B1310$+$18  &33.163 & M53    &18.9 & 55.80&84.20 & 0.002  & 0.3& 
\cite{kapw91}\\
B1516$+$02A & 5.554 & M5     & 7.0 &                &    &   &    & 
\cite{awkp97}\\
B1516$+$02B & 7.947 & M5     & 7.0 & 6.86 & 3.04 &0.14       &    &
\cite{awkp97}\\
B1620$-$26  & 11.076& M4     & 1.8 & 91.44& 64.81&0.03   & 0.3&
\cite{lbb+88,tat93,bfs93,tacl99}\\
B1639$+$36A & 10.378& M13    & 7.7 &                &    &   &    &
\cite{kapw91}\\
B1639$+$36B & 3.528 & M13    & 7.7 & 1.26 & 1.39 & 0.005     & 0.2&
\cite{kapw91}\\
J1701$-$30  & 5.242 & NGC6266& 6.7 & 3.81 & 3.48 & $>0.0$       & 0.2&
\cite{dlm+01}\\
B1718$-$19  &1004.037&NGC6342& 7.0 & 0.26 & 0.35 & $>0.0$       & 0.1&
\cite{lbhb93}\\
J1740$-$53  & 3.650 & NGC6397& 2.2 & 1.35 & 1.66 & $>0.0$       & 0.2&
\cite{dlm+01}\\
B1744$-$24A & 11.563& Terzan5& 7.1 & 0.08 & 0.12 & $>0.0$       & 0.1& 
\cite{lmd+90,nt92}\\
J1748$-$2446C& 8.436& Terzan5& 7.1 &                &    &   &    &
\cite{lmbm00}\\
B1745$-$20  &288.603&NGC6440 & 5.8 &                &    &   &    &
\cite{lmd96}\\
B1802$-$07  & 23.101& NGC6539& 3.1 & 2.62 & 3.92 & 0.21  & 0.3&
\cite{dbl+93,tamt93}\\
J1807$-$24  & 3.059 & NGC6544& 2.5 & 0.071& 0.012& $>0.0$       &0.009&
\cite{rgh+01,dlm+01}\\
B1820$-$30A & 5.440 & NGC6624& 8.0 &                &    &   &    & 
\cite{bbl+94}\\
B1820$-$30B &378.596& NGC6624& 8.0 &                &    &   &    &
\cite{bbl+94,lmd96}\\
B1821$-$24  & 3.054 & M28    & 5.5 &                &    &   &    &
\cite{lbm+87}\\
J1910$-$59  & 3.266 & NGC6752& 3.9 & 0.865 & 1.27& $>0.0$       &0.19&
\cite{dlm+01}\\
J1910$+$0004& 3.619 & NGC6760& 4.1 & 0.14  & 0.04& $>0.0$       & 0.02&
\cite{dma+93}\\
B2127$+$11A &110.665& M15    &10.0 &                &    &   &    &
\cite{wkm+89,and92}\\
B2127$+$11B & 56.133& M15    &10.0 &                &    &   &    &
\cite{agk+90,and92}\\
B2127$+$11C & 30.529& M15    &10.0 & 0.34  & 2.52&0.68   & 0.9&
\cite{agk+90,and92}\\
B2127$+$11D & 4.803 & M15    &10.0 &                &    &   &    &
\cite{pakw91,and92}\\
B2127$+$11E & 4.651 & M15    &10.0 &                &    &   &    &
\cite{pakw91,and92}\\
B2127$+$11F & 4.027 & M15    &10.0 &                &    &   &    &
\cite{and92}\\
B2127$+$11G & 37.660& M15    &10.0 &                &    &   &    &
\cite{and92}\\
B2127$+$11H & 6.743 & M15    &10.0 &                &    &   &    &
\cite{and92}\\
\hline
\end{tabular}
\caption{\sl
Parameters for the 47 pulsars currently known in globular clusters.
Listed are the spin period $P$, cluster name and
the distance $d$ to the cluster. For binary pulsars we also list
the binary period $P_b$, the projected semi-major axis of the orbit $x$
in units of light seconds, the orbital eccentricity $e$ and the 
companion mass $m_2$ evaluated from the mass function assuming
a pulsar mass of 1.4 M$_{\odot}$ and an inclination angle of 60 degrees
(\S \ref{sec:tbin}) or (when known) from independent measurements. 
Key publications for each pulsar are referenced to the bibliography.
}
\label{tab:gcpsrs}
\end{table}

\end{document}